\documentclass[journal,onecolumn]{IEEEtran}
\usepackage{cite}
\usepackage{commath}
\usepackage{amsmath,amssymb,amsfonts}
\usepackage{scalefnt}
\usepackage{algorithmic}
\usepackage{graphicx}
\usepackage{textcomp}
\usepackage{nicefrac}
\usepackage{xcolor}
\usepackage{steinmetz}
\usepackage{multicol}
\usepackage{lipsum}
\usepackage{enumitem}  
\usepackage{float}
\usepackage{makecell}
\usepackage{longtable}
\usepackage{subfigure}
\usepackage{cuted} 
\usepackage{dblfloatfix}    
\usepackage{tikz}
\usetikzlibrary{shapes.geometric, arrows}
\usetikzlibrary{positioning,decorations.pathreplacing}
 \tikzstyle{arrow}=[draw, -latex]
\tikzstyle{line}=[draw] 

\tikzstyle{startstop} = [rectangle, rounded corners, minimum width=3cm, minimum height=0.5cm, text centered, draw = black, fill = white]
\tikzstyle{io} = [trapezium, trapezium left angle = 70, trapezium right angle = 110, minimum width = 0.1cm, minimum height=0.5cm, text centered, draw = black, fill = white]
\tikzstyle{process} = [rectangle, minimum width=0.1cm, minimum height=0.5cm, text centered, draw=black, fill = white]
\tikzstyle{decision} = [diamond, aspect=2, text width=10em, inner sep=-5pt, text centered, draw=black, fill = white]
\tikzstyle{arrows} = [thick, ->, >=stealth]

\makeatletter
\let\latexlabel\ltx@label
\makeatother

\renewcommand{\IEEEbibitemsep}{0pt plus 0.5pt}
\makeatletter
\IEEEtriggercmd{\reset@font\normalfont\fontsize{7.9pt}{8.40pt}\selectfont}
\makeatother
\IEEEtriggeratref{1}

\begin{document}
\definecolor{dgreen}{rgb}{0.0, 0.5, 0.0}
\setlength{\abovedisplayskip}{3pt}
\setlength{\belowdisplayskip}{3pt}

\title{\vspace{-0.75cm} Stability assessment for multi-infeed grid-connected VSCs modelled in the admittance matrix form}

\author{Luis Orellana, Luis Sainz, Eduardo Prieto-Araujo,~\IEEEmembership{Member,~IEEE,} and Oriol Gomis-Bellmunt,~\IEEEmembership{Fellow Member,~IEEE}\thanks{

This project has received funding from the European Union's Horizon 2020 research and innovation programme under the Marie Sklodowska-Curie grant agreement no. 765585. This document reflects only the author’s views; the European Commission is not responsible for any use that may be made of the information it contains. This work has also been partially funded by FEDER/Ministerio de Ciencia, Innovaci\'on y Universidades - Agencia Estatal de Investigaci\'on, Project RTI2018-095429-B-I00 and by the ICREA Academia program.

L. Orellana, E. Prieto-Araujo, and O. Gomis-Bellmunt are with the Centre d'Innovaci\'o Tecnol\`ogica en Convertidors Est\`atics i Accionaments, Departament d'Enginyeria El\`ectrica, Universitat Polit\`ecnica de Catalunya, Barcelona 08028, Spain (e-mail: luis.orellana@upc.edu; eduardo.prieto-araujo@upc.edu; oriol.gomis@upc.edu). E Prieto-Araujo is also a Serra H\'unter lecturer. O Gomis-Bellmunt is also an ICREA academia researcher.

L. Sainz is with Department of Electric Engineering, ETS d’Enginyeria Industrial
de Barcelona, Universitat Politecnica de Catalunya, Barcelona 08028,
Spain (e-mail: luis.sainz@upc.edu).

}}


\maketitle
\begin{abstract}

    The increasing use of power electronics converters to integrate renewable energy sources has been subject of concern due to the resonance oscillatory phenomena caused by their interaction with poorly damped AC networks. Early studies are focused to assess the controller influence of a single converter connected to simple networks, and they are no longer representative for existing systems. Lately, studies of multi-infeed grid-connected converters are of particular interest, and their main aim is to apply traditional criteria and identify their difficulties in the stability assessment. An extension of traditional criteria is commonly proposed as a result of these analysis, but they can be burdensome for large and complex power systems. The present work addresses this issue by proposing a simple criterion to assess the stability of large power systems with high-penetration of power converters. The criterion has its origin in the mode analysis and positive-net damping stability criteria, and it addresses the stability in the frequency domain by studying the eigenvalues magnitude and real component of dynamic models in the admittance matrix form. Its effectiveness is tested in two case studies developed in Matlab/Simulink which compare it with traditionally criteria, proving its simplicity.
    
\end{abstract}


\IEEEpeerreviewmaketitle


\section{Introduction}

    \IEEEPARstart{T}{\lowercase{h}}e use of grid-connected power converters has been increasing due to the need to connect large renewable energy resources to the  AC power network. These resources are typically connected to the AC grid by means of voltage source converter technology (VSC) which play an important role in the transmission system development. However, VSCs also bring new challenges and problems due to the interaction with components of the traditional power system such as synchronous generators, power transformers and transmission lines. One of the most important problems is the oscillatory phenomena caused by the interaction between the VSC control and the grid. These oscillations can lead to instabilities specially in poorly damped networks~\cite{Revel2014DynamicsServices,Buchhagen2015BorWin1Grids,Sun2017RenewableOpportunities}. There are non-damped cases when the system maintains a sustained oscillation due to non-linearities such as saturation and limiters~\cite{Shah2018Large-signalConverters,Wu2020InclusionPMSGs}.
        
    The commonly used methods to model grid-connected VSC systems to study the oscillatory phenomena are the state-space and impedance-based modelling approaches~\cite{Coelho2002Small-signalSystems,Harnefors2007}. The first one represents the system as a set of linear equations in the time domain, but it requires detailed information of the control code which is possibly not available. On the other hand, impedance-based modelling approach is based on the impedance characterization of the system (e.g., detailed knowledge about the converts is not needed) which can be expressed as a transfer function in the \textit{s}-domain. Stability criteria such as the Nyquist stability criterion~\cite{Harnefors2007ModelingMatrices,Harnefors2008Frequency-domainDesign}, impedance-based stability criterion~\cite{Sun2011Impedance-BasedInverters,Cespedes2014ImpedanceConverters,Zhang2015Impedance-basedSystems}, and the positive-net damping stability criterion (PND)~\cite{Sainz2017Positive-Net-DampingSystems,Cheah-Mane2017} have been used to study the controller influence of a single VSC over the stability of simple networks in the frequency domain. 
    
    Stability studies of multi-infeed VSC-based AC grids are currently of great interest and different approaches to assess stability of the nodal admittance matrix in the $s$-domain by using modal analysis are presented in\cite{Ebrahimzadeh2018HarmonicPlants,Zhan2019Frequency-domainRenewables,Xing2020ResonanceGenerators}, and in the frequency domain based on the GNC in~\cite{Wang2014ModelingSystem,Pedra2020Three-PortGrids,Li2021StabilityMatrix}. These studies use the nodal admittance matrix modelling approach, an enhancement of impedance-based modelling methods, as it is simple and powerful when characterizing multi-infeed large power systems. In~\cite{Ebrahimzadeh2018HarmonicPlants}, the stability of MIMO systems is assessed by looking at the nodal admittance matrix poles; the contribution in~\cite{Zhan2019Frequency-domainRenewables} studies the zeros of the nodal admittance or loop impedance matrix determinant; the stability is assessed with the damping coefficient and the negative-resistance effect of the resonance modes in~\cite{Xing2020ResonanceGenerators}; however, all mentioned studies are conducted in the $s$-domain, and  there is a preference in the industry to work with measurements in the frequency domain as it allows the use of black-box models.
    
    Typically, the stability of the nodal admittance matrix is assessed in the frequency domain with the generalized Nyquist criterion (GNC) which extends the Nyquist criterion for single-input and single-output (SISO) to multiple-input and multiple-output (MIMO) dynamic systems as introduced in~\cite{Harnefors2007ModelingMatrices}. For example, the stability of MIMO systems is assessed with the GNC by using the impedance-based approach of the nodal admittance matrix form for a  three-phase meshed and balance power system in~\cite{Wang2014ModelingSystem}, for hybrid AC/DC grids~\cite{Pedra2020Three-PortGrids}, and for large-scale multiconverter systems in~\cite{Li2021StabilityMatrix}. However, studies in~\cite{Zhang2020Impedance-BasedCriteria,Liao2020Impedance-BasedPoles} show that GNC may lead to a wrong stability conclusion due to misleading associations at the time of deriving the closed-loop transfer function. Further drawbacks using GNC were identified during the present study such as computational efforts (i.e., time and memory) at the time of evaluating the open-loop in the [$-j\infty,+j\infty$] frequency range in order to contour the unstable poles located in the right half-plane (RHP) for high-order admittance matrices. Additionally, the analysis of large networks is challenging due to the numerous Nyquist curves of eigenvalues.
    
    The resonance mode analysis (RMA), introduced in \cite{Xu2005HarmonicAnalysis}, helps to identify harmonic resonance modes for systems in the admittance matrix form; however, no stability criterion is proposed. The letter in~\cite{Chou2020Frequency-DomainSystems} uses the "peak-picking" method and the "circle fit" method introduced in~\cite{Zhi-Fang2001ModalAnalysis} to analyse these resonance modes obtained from measurement data, yet no contribution about addressing the stability beyond the traditional criteria was made.
    
    To address the above concerns in traditional frequency domain stability criteria, this paper contributes with a new stability criterion, called as positive-mode damping (PMD) stability criterion, which is based on the RMA and the PND stability criteria. The features of the proposed criterion are summarized as follows:
    
     \begin{enumerate}
        \item The oscillatory modes can be characterized in the frequency domain; 
        \item it does not require detailed information;
        \item the system stability can be assessed from experimental measurements (e.g., black-box models);
        \item it is not affected by aggregation of system elements;
        \item it is easy to use;
        \item it requires less effort to calculate and evaluate than traditional stability criteria;
        \item its application can be programmed.
    \end{enumerate}
   
    The effectiveness of proposed stability criterion is tested in two case studies implemented in Matlab/Simulink, comparing its results with the corresponding eigenvalue analysis and the GNC criterion application over the same system. The first case study is built to demonstrate that the proposed criterion provides a correct stability assessment while others methods fail. It is composed by three grouping options, where each one studies the closed-loop stability of the system, in the admittance matrix form, by diving it into two subsystems. The instability condition is the same for all grouping options, but the approach to build each of the subsystems is different between them, leading to possible wrong stability assessment conclusions. The second case has two examples in order to show the simplicity of the proposed method for assessing the stability of large and complex networks. The visualisation of results and the computation effort for a large network is compared between the proposed criterion and commonly used closed-loop stability criteria in the frequency domain.  
    

\section{Grid-connected VSC modelling}

\begin{figure}[htbp]
    \subfigure[]{\includegraphics[width=0.5\linewidth]{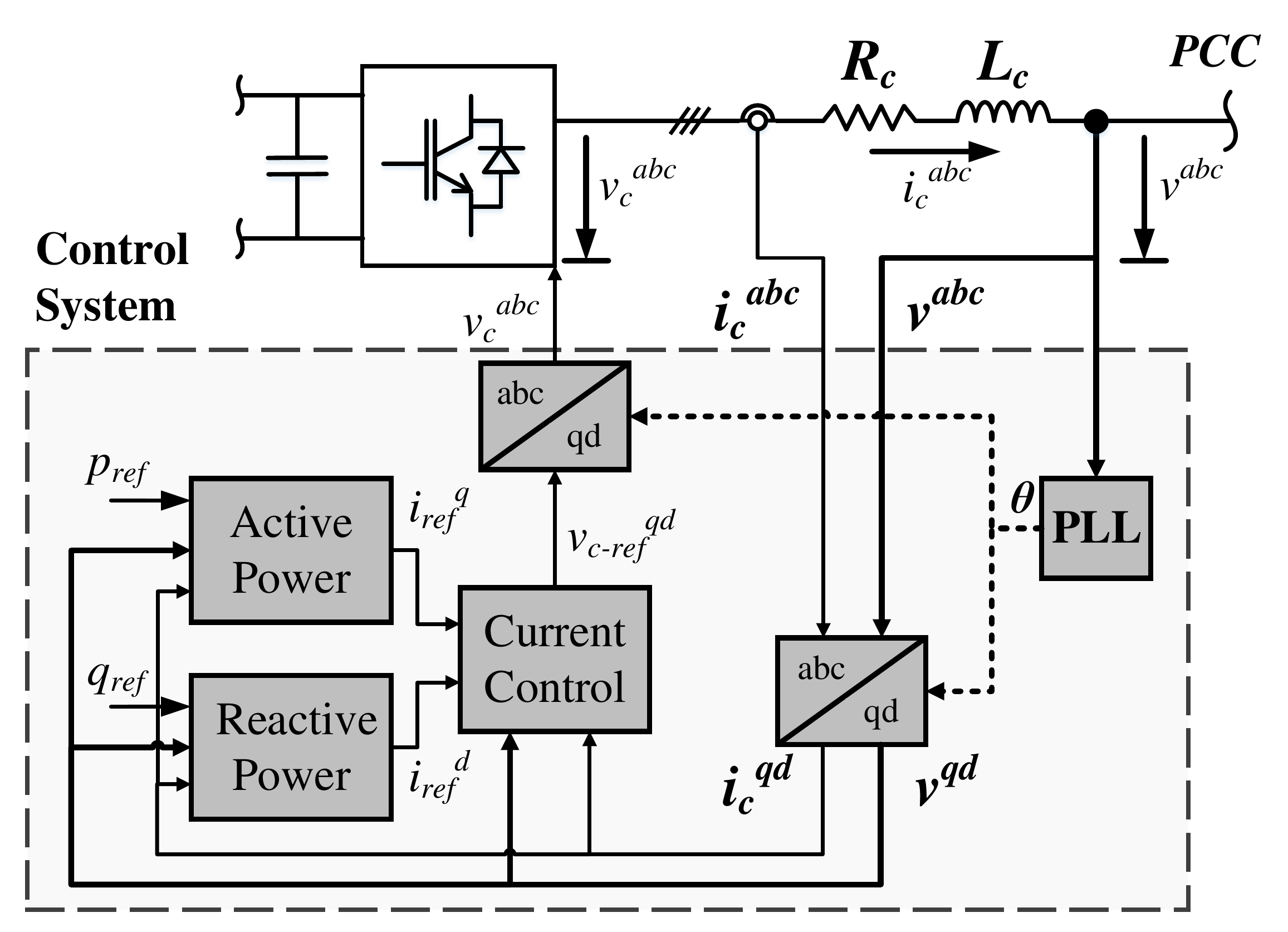}\label{subfig:T_d}}
    \subfigure[]{\includegraphics[width=0.5\linewidth]{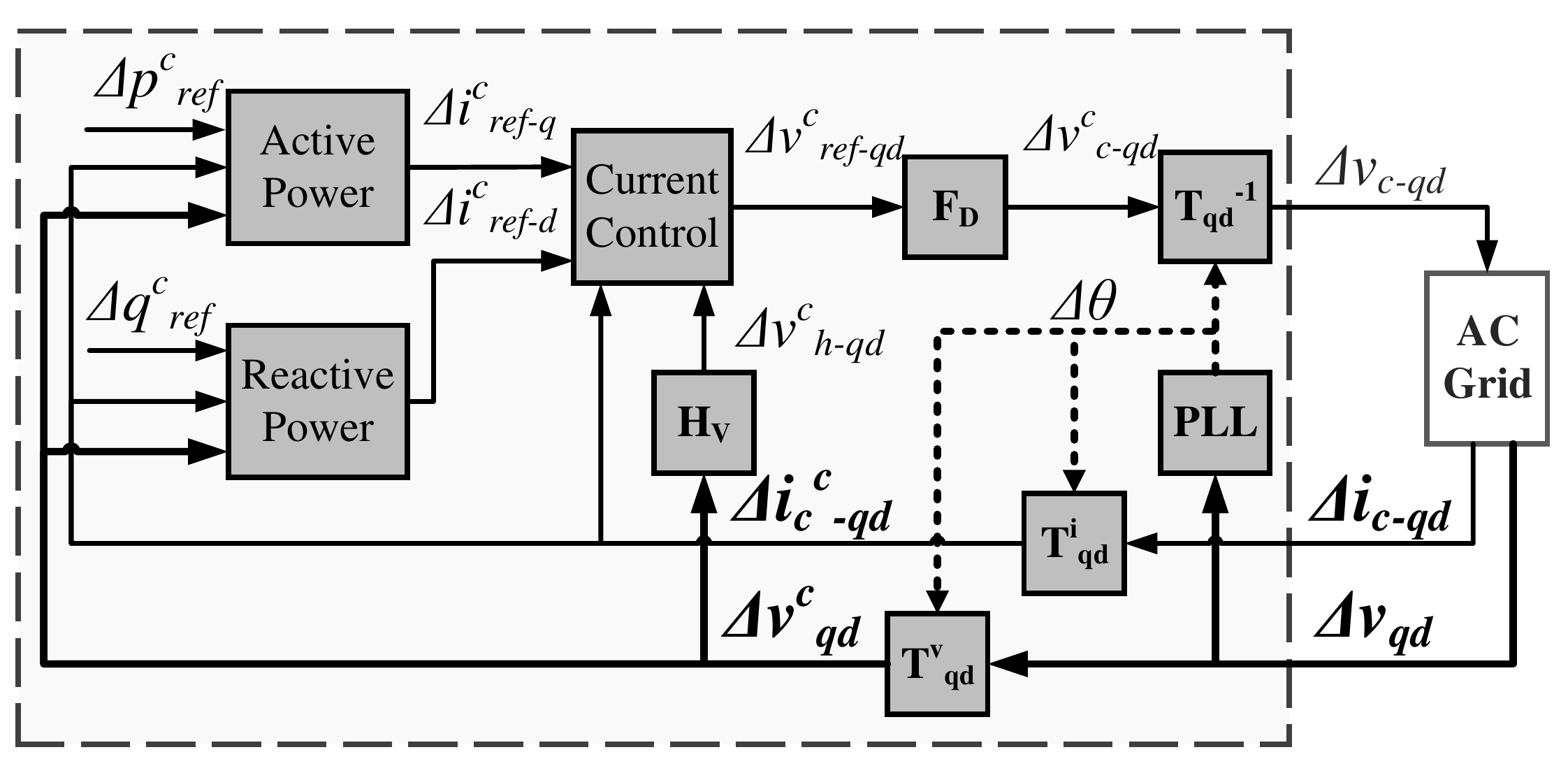}\label{subfig:T_cs}}
    \caption{Grid-connected VSC. (a) Schematic diagram control structure. (b) Block diagram small-signal model.}
    \end{figure}

    Fig.~\ref{subfig:T_d} displays the control structure of a generic grid-connected VSC. The model is an averaged three-phase converter, which uses vector control strategy with a cascaded controller to control active and reactive power~\cite{Xu2007DirectConverters}.
    
    The dynamics of the VSC can be modelled by both state-space and impedance-based modelling approaches~\cite{Coelho2002Small-signalSystems,Harnefors2007}. The small-signal model for stability studies at the point of common coupling (PCC), as displayed in Fig.~\ref{subfig:T_d}, can be formulated to state-space equations as\begin{eqnarray}
    \dot{x}(t) & =\mathbf{A}x(t)+\mathbf{B}u(t) & u(t)=\Delta v_{qd}(t)\notag\\
    y(t) & =\mathbf{C}x(t)+\mathbf{D}u(t) & y(t)=\Delta i_{c-qd}(t),\label{eq:T_ss}
    \end{eqnarray} 
    
    \noindent where $x(t), u(t)$ and $y(t)$ are the states, input and output of the system state-space representation; and by a two-by-two impedance matrix, where each of its elements is a transfer function in the \textit{s}-domain as follows \begin{eqnarray}
    \Delta v_{qd}=\underbrace{\begin{bmatrix}
    Z_{vsc-qq}(s) & Z_{vsc-qd}(s) \\
    Z_{vsc-dq}(s) & Z_{vsc-dd}(s)
    \end{bmatrix}}_{\mathbf{Z_{vsc}}(s)}\Delta i_{c-qd},
    \end{eqnarray} 

    \noindent where $\Delta v_{qd} = [\Delta v_q \enspace \Delta v_d]^T $ and $\Delta i_{c-qd} = [\Delta i_{c-q} \enspace \Delta i_{c-d}]^T$. A comparison between both small-signal modelling approaches in a local reference is described in \cite{Orellana2019OnGrids} for further information.
    
    When two or more VSCs are connected to the AC grid, they cannot longer be in a local reference. In other words, they all should be referenced to a reference or slack bus in the AC network, as detailed in~\cite{Rygg2017AAnalysis} (see the Appendix for more details). 
    
    In the impedance-modelling approach, the converter impedance can be easily added to the network nodal admittance matrix by means of its admittance $\mathbf{Y_{vsc}}(s)$=$(\mathbf{Z_{vsc}}(s))^{-1}$ as other $\mathbf{Y_{RL}}$ series or $\mathbf{Y_{C}}$ shunt connected elements by applying the voltage node method, \begin{eqnarray}
    \mathbf{Y_{RL}} (s) = \begin{bmatrix}  R+L s & \omega L \\  -\omega L & R+L s \end{bmatrix}^{-1}  \mathbf{Y_C} (s)  = \begin{bmatrix}C s & C \omega \\ -C \omega & C s \end{bmatrix}. \label{eq:T_z}
    \end{eqnarray}
    

\section{Multi-infeed grid-connected VSCs modelling}  \label{sec:Multi-infeed}

\begin{figure}[htbp]
    \centering
    \includegraphics[width=0.5\linewidth]{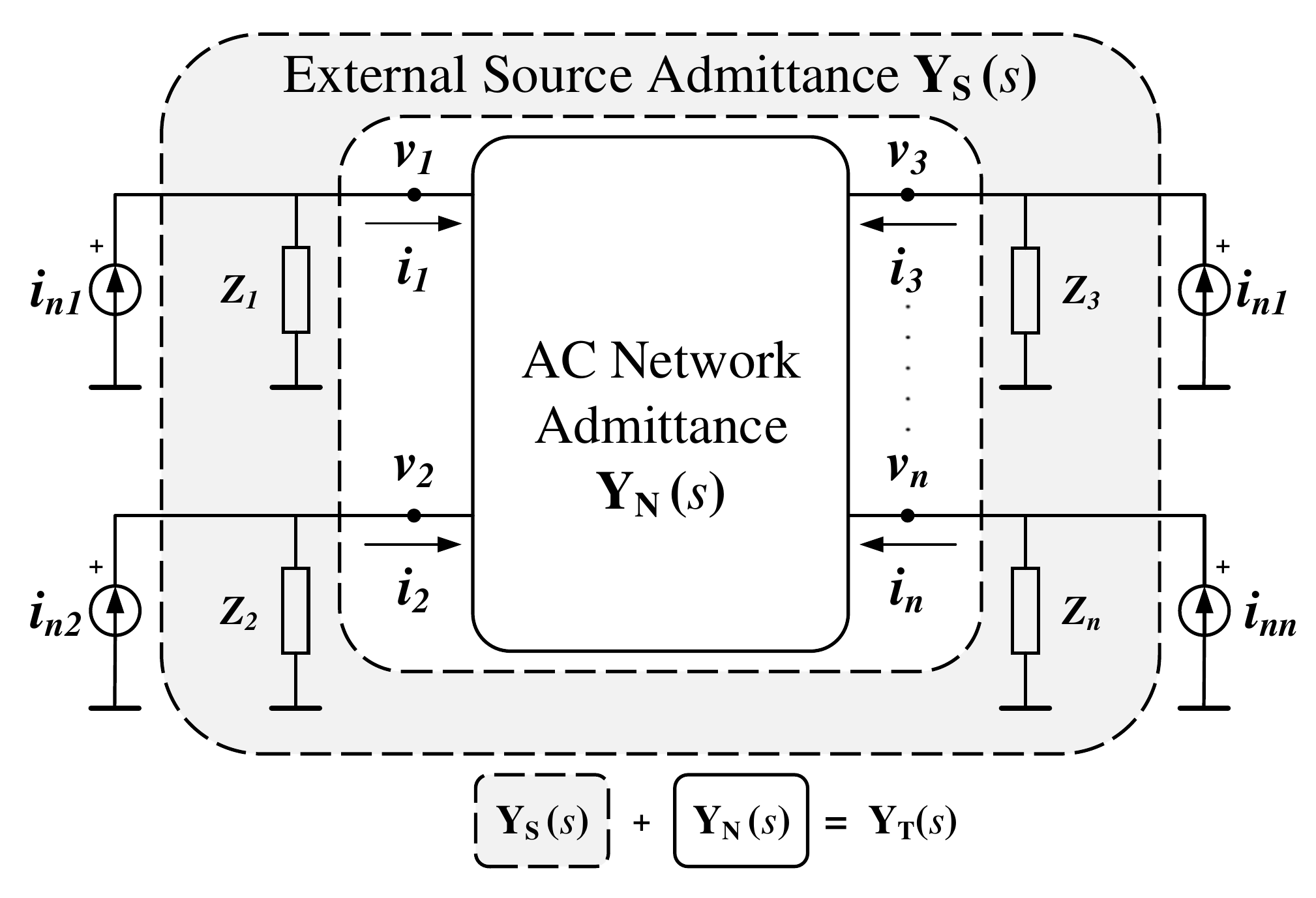}
    \caption{Schematic diagram of multi-infeed grid connected VSCs.}
    \label{fig:T_mimo}
    \end{figure}

    Fig.~\ref{fig:T_mimo} shows a schematic diagram used to represent multi-infeed VSC-based AC grids as carried out in~\cite{Amin2017Small-SignalMethods,Liu2018AnRenewables,Liu2018ImpedanceSystems,Ebrahimzadeh2018HarmonicPlants,Pedra2020Three-PortGrids,Zhang2020Impedance-BasedCriteria,Li2021StabilityMatrix}. The network is characterized by its admittance matrix $\mathbf{Y_N}(s)$, and the voltages and currents at its terminals (i.e., $n$ the number of buses) are $v=[v_1 \enspace...\enspace v_n]^T$ and $i=[i_1 \enspace ...\enspace i_n]^T$. The external elements (e.g., VSC converters) connected at the network buses are represented by their Norton equivalent circuits formed by the Norton currents $i_n=[i_{n1} \enspace....\enspace i_{nn}]^T$ with their corresponding impedance connected in parallel.

    The relation between voltages and currents at the AC grid terminals are expressed as \begin{eqnarray}
    \left.
        \begin{array}{ll}
        i = \mathbf{Y_N}(s) v\\
        i = i_n-\mathbf{Y_S}(s) v
        \end{array}
    \right\}  i_n= (\mathbf{Y_N}(s)+\mathbf{Y_S}(s))v=\mathbf{Y_T}(s)v,
    \label{eq:T_znm}
    \end{eqnarray}

    \noindent where $\mathbf{Y_T}(s)$ and $\mathbf{Y_S}(s)$ are the nodal admittance matrix transfer functions of the full system and the external elements' impedance which can be easily obtained by applying the voltage node method.

    
\section{Stability assessment criteria}

    A small-signal stability analysis can be performed using both state-space and impedance-based dynamic models. In state-space representations, stability is commonly assessed by using eigenvalue analysis. On the other hand, the stability in impedance-based models is typically assessed with closed-loop stability criteria, that studies the open-loop formed by the ratio of the two subsystems' impedance partitioned accordingly to each criterion~\cite{Sun2011Impedance-BasedInverters,Cespedes2014ImpedanceConverters}. The stability of MIMO systems is typically assessed applying the GNC to the open-loop~\cite{Harnefors2007ModelingMatrices}.
    
    \subsection{Eigenvalue analysis} \label{criterion1} 
    
    Stability of the multi-infeed grid-connected VSCs in Fig.~\ref{fig:T_mimo} can be studied from state-space equations in~\eqref{eq:T_ss}, where $u(t)=i_n(t)$ and $y(t)=v(t)$ are the input and output variables of the system state-space. Stability can be addressed by obtaining the eigenvalues $\lambda_i = \sigma_i \pm j \omega_i$ of the state-space matrix calculated as $\abs{\mathbf{A}-\lambda \mathbf{I}}=0$. The stability criterion is built on the real and imaginary part: the real part $\sigma_i=\textrm{Re}\{\lambda_i\}$ represents the system damping and the imaginary $\omega_i=\textrm{Im}{\{\lambda_i\}}$ the frequency of oscillation. The system is unstable if it contains any eigenvalue in the RHP (i.e., $\sigma_0>0$ means oscillatory instability of the frequency mode $\omega_0$)~\cite{Kundur1994PowerControl}. Considering~\eqref{eq:T_znm}, the expression in~\eqref{eq:T_ss} can be rearranged as \begin{eqnarray}
    v = \underbrace{[\mathbf{C}(s \mathbf{I}-\mathbf{A})^{-1}\mathbf{B}+\mathbf{D}]}_{\mathbf{Z_T}(s)} i_n,
    \label{eq:T_zss}
    \end{eqnarray} 
    
    \noindent where $\mathbf{Z_T}(s)=\mathbf{Y_T}(s)^{-1}$ is the impedance matrix transfer function which can be expressed from~\eqref{eq:T_zss} in the following form \begin{eqnarray}
    \mathbf{Z_{T}}(s)  = \dfrac{1}{\abs{s \mathbf{I}-A}}\mathbf{C}[\textrm{adj}(s \mathbf{I}-\mathbf{A})]\mathbf{B}+\mathbf{D} = \dfrac{1}{D(s)}\mathbf{Z_{Tb}}(s), 
    \label{eq:T_ztf}
    \end{eqnarray} 
    
    \noindent being adj$(s\mathbf{I}-\mathbf{A})$ the adjoint matrix of $(s\mathbf{I}-\mathbf{A}$) and $\mathbf{Z_{Tb}}(s)$ the adjoint matrix of $\mathbf{Z_{T}}(s)$.
    
    It can be noted from \eqref{eq:T_ztf} that the poles of $\mathbf{Z_T}(s)$ are the roots of the denominator $D(s)=\abs{s\mathbf{I}-\mathbf{A}}$, namely the eigenvalues $\lambda_i$ of the state-space matrix $\mathbf{A}$~\cite{maciejowski1989multivariable, chen1999linear}. Therefore, system stability can be assessed by either the eigenvalues $\lambda_i$ of $\mathbf{A}$ as well as the poles of $\mathbf{Z_T}(s)$.
    
    Eigenvalue analysis is a simple stability criterion which helps to analyse the stability of large systems in a short time. However, it requires detailed information in order to model real systems, which sometimes is not available (e.g., control structure and parameters of power converters).
    
    \subsection{Generalized Nyquist criterion} The expression in~\eqref{eq:T_znm} can be rewritten as \begin{eqnarray}
    v=(\mathbf{I}+\mathbf{L}(s))^{-1}\mathbf{Z_N}(s) i_n &  \mathbf{L}(s)=\mathbf{Z_N}(s) \mathbf{Y_S}(s),
    \label{eq:T_ztii}
    \end{eqnarray}
    
    \noindent where $\mathbf{Z_N}(s) = (\mathbf{Y_N}(s))^{–1}$ and $\mathbf{I}$ is an $n$th order identity matrix. 
    
    If the open-loop $\mathbf{L}(s)$ does not have any RHP poles, the stability of the closed-loop system in~\eqref{eq:T_ztii}  can be assessed by the GNC, which extends the traditional Nyquist criterion for SISO systems to Nyquist curves of the eigenvalues of $\mathbf{L}(s)$~\cite{Harnefors2007ModelingMatrices,Amin2017Small-SignalMethods}. This means that system stability is assessed by counting the clockwise encirclements of the eigenvalues $\lambda_{ni}$ of $\mathbf{L}(s)$ around the (–1, $j$0) point. This is valid for multi-infeed grid-connected VSCs modelling in~\eqref{eq:T_ztii} because the network $\mathbf{Z_N}(s)$ is passive and the external components in $\mathbf{Y_S}(s)$ are individual subsystems, which in stand-alone operation are stable (i.e., $\mathbf{Z_N}(s)$ and $\mathbf{Y_S}(s)$ do not have any RHP poles). However, this might not be valid anymore if some of these individual subsystems come from an aggregation of part of the network containing VSCs. This aggregation could be unstable due to the interaction of the grouped VSCs and network passive components, introducing RHP poles in $\mathbf{Y_S}(s)$~\cite{Zhang2020Impedance-BasedCriteria}.
    
    On the other hand, part of the drawbacks identified in eigenvalue analysis, such as the required knowledge of the complete control structure, can be solved by using a GNC-based stability analysis. Stability can be assessed from frequency dependent models provided by manufactures (e.g., converter impedance curves) which are obtained from numeric simulation or experimental measurements. However, applying GNC, wrong stability conclusions might be made due to order-cancellation, open-loop RHP poles, and improper minor-loop gain or impedance ratio has been identified in~\cite{Zhang2020Impedance-BasedCriteria,Liao2020Impedance-BasedPoles}.
    
   
\section{Positive mode damping criterion} 
    
    It is well-known that instabilities are related to low-damped network resonances. This has been proved for a single grid-connected VSC with the PND stability criterion in~\cite{Harnefors2007ModelingMatrices,Sainz2017Positive-Net-DampingSystems,Cheah-Mane2017} which evaluates the damping of the SISO transfer function $Z_T(s)$ at resonance frequencies. It is stated that a  system is stable if and only if the damping is positive at these resonance frequencies, i.e., $\textrm{Re}\{Z_T(j\omega_x)\} > 0$. It is also worth mentioning that the PND stability criterion in~\cite{Sainz2017Positive-Net-DampingSystems} evaluates the closed-loop function (i.e., $Z_T(j\omega_x)$) to address the stability; therefore, it is not affected by misleading associations of the system elements as it is the case of the Nyquist criterion, which evaluates the open-loop function~\cite{Zhang2020Impedance-BasedCriteria,Liao2020Impedance-BasedPoles}. The proposed stability approach, called positive-mode damping (PMD) stability criterion, extends the PND stability criterion to multi-infeed grid-connected VSCs by means of the RMA.
    
    The RMA provides an effective tool for evaluating the resonances of networks modelled in the admittance matrix form~\cite{Xu2005HarmonicAnalysis}, by addressing the statement in~\eqref{eq:T_znm} as \begin{eqnarray}
    v= \mathbf{Y_T}(j \omega_x) i_n  &  \mathbf{Y_T}(j \omega_x)=\mathbf{L} \mathbf{\Lambda_Y} \mathbf{T}, \label{eq:T_zde}
    \end{eqnarray}
    
    \noindent where $\mathbf{Y_T}(j\omega_x)$ is the system admittance matrix at frequency $\omega_x$; $v$ and $i_n$ are the nodal voltage and current injection vectors; $\mathbf{L}$ and $\mathbf{T}$ are the right and left eigenvector matrices; and $\mathbf{\Lambda_Y}$ is the diagonal eigenvalue matrix, \begin{eqnarray}
    \mathbf{\Lambda_Y} = 
    \begin{bmatrix}
    \lambda_{y1}   & 0      & . & . & . & 0 \\
           0 & \lambda_{y2} & . & . & . & 0 \\
          . &  .        & . &   &   & . \\
           . &  .        &   & . &   & . \\
           . &  .        &   &   & . & . \\
           0 & 0         & . & . & . & \lambda_{yn}
    \end{bmatrix}
    . \label{eq:T_lay} 
    \end{eqnarray}
    
    It must be noted that the inverse diagonal eigenvalue matrix $\mathbf{\Lambda_Y}$ is the diagonal eigenvalue matrix of $\mathbf{Z_T}(j\omega_x)$ in \eqref{eq:T_zss}, \begin{eqnarray}
    \mathbf{Z_T}(j\omega_x) = (\mathbf{Y_T}(j\omega_x))^{-1}=\mathbf{L} \mathbf{\Lambda_Z} \mathbf{T} \notag \\
    \mathbf{\Lambda_Z} =  [\lambda_{z1} \enspace \lambda_{z2} \enspace ... \enspace \lambda_{zn}] \mathbf{I} \qquad
    \qquad \lambda_{zi} =\dfrac{1}{\lambda_{yi}},
    \label{eq:T_lazss}
    \end{eqnarray}
    
    \noindent where the diagonal terms of $\mathbf{\Lambda_Z}$ are called modal impedances $\lambda_{zi}$.
    
    Parallel resonance phenomena is associated with the singularity of $\mathbf{Y_T}(j\omega)$ which happens when one of its eigenvalues $\lambda_{yi}$ approaches 0. The resonance modes can also be identified from peaks values at the magnitude modal impedance $\abs{ \lambda_{zi}}$ curves in the frequency domain~\cite{Xu2005HarmonicAnalysis}.
    
    It must be noted that the poles of $\mathbf{Z_T}(s)$ (i.e., the eigenvalues of the state-space matrix A) are the same as the poles of the modal impedances of $\mathbf{\Lambda_Z}(s)$, \begin{eqnarray}
    \mathbf{\Lambda_Z}(s) & = & \dfrac{1}{D(s)}\mathbf{T} \mathbf{Z_{Tb}}(s) \mathbf{L}=\dfrac{1}{D(s)} \mathbf{\Lambda_{Zb}}(s) \notag \\
    \mathbf{\Lambda_{Zb}}(s) & = & \mathbf{T} \mathbf{Z_{Tb}}(s) \mathbf{L} = 
    [\lambda_{zb1} \enspace \lambda_{zb2} \enspace ... \enspace \lambda_{zbn}] \mathbf{I}.
    \label{eq:T_la} 
    \label{eq:T_laztf}
    \end{eqnarray}
    
    Therefore, the stability of the system in~\eqref{eq:T_lazss} can be assessed with the diagonal matrix $\mathbf{\Lambda_Z}$, and the analysis can be carried out independently for each modal impedance $\lambda_{zi}$ as a SISO system by applying the PND stability criterion to each $\lambda_{zi}$~\cite{Sainz2017Positive-Net-DampingSystems}. These modal impedances can be expressed as,     \begin{eqnarray}
    \lambda_{zi}(j\omega_x) =\dfrac{\lambda_{zbi}(j\omega_x)}{\underset{i=1}{\overset{i_p}{\prod}}(j\omega_x - p_i)(j\omega_x - p_i^*)} = \dfrac{G(j\omega_x)}{(j\omega_x - p_0)(j\omega_x - p_0^*)} = \dfrac{G(j\omega_x)}{\sigma_0^2+\omega_0^2-\omega_x^2-j 2\sigma_0 \omega_x},
    \label{eq:T_lazi} 
    \end{eqnarray}  
    
    \noindent where $p_0=\sigma_0\pm j\omega_0$ is pair of complex conjugate poles of $\lambda_{zi}$ corresponding to a certain system oscillatory mode which match with eigenavalues $\lambda_0$ of the state-space matrix $\mathbf{A}$, and $G(j\omega_x)$ is a polynomial expression representing the rest of the terms of $\lambda_{zi}$. 
    
    It can be observed that the modal impedance in \eqref{eq:T_lazi} will be maximum or have a peak value at the oscillation frequency (i.e., $\omega_x\approx\omega_0$) in the case of a poorly damped oscillatory mode (i.e., $\abs{\sigma_0}<<\abs{\omega_0}$), which is the main concern in academia and industry due to the following reasons: (a) a power system maintains stable operation for strongly damped modes with large negative $\sigma_i$; and (b) monotonic instability caused by large positive $\sigma_i$ occurs less often in power systems. In these cases, the growing oscillations caused by large positive $\sigma_0$ are sustained due to saturation and limiters non-linearities~\cite{Shah2018Large-signalConverters,Wu2020InclusionPMSGs}. 

    If $\omega_x$ is within the small neighbourhood of $\omega_0$, $G(j\omega_x) \approx G(j\omega_0)=G_r+j G_x$ where $G_r$ and $G_x$ are constant complex numbers dependent on $\omega_0$~\cite{Liu2018AnRenewables}. Thus, $\lambda_{zi}$ can be further expressed as \begin{eqnarray}
    \lambda_{zi}(j\omega_x) = \dfrac{\sigma_0^2+\omega_0^2-\omega_x^2+j2\omega_x\sigma_0^2}{(\sigma_0^2+\omega_0^2-\omega_x^2)^2+(2\sigma_0 \omega_x)^2} (G_r+jG_x) = 
    \lambda_{zi,r}(\omega_x)+j\lambda_{zi,x}(\omega_x)
    \label{eq:T_law}
    \end{eqnarray}
    
    \noindent where \begin{eqnarray}
    \lambda_{zi,r}(\omega_x) =    \dfrac{(\sigma_0^2+\omega_0^2-\omega_x^2)G_r-2\omega_x\sigma_0^2 G_x}{(\sigma_0^2+\omega_0^2-\omega_x^2)^2+(2\sigma_0 \omega_x)^2} \notag \\
    \lambda_{zi,x}(\omega_x) = \dfrac{(\sigma_0^2+\omega_0^2-\omega_x^2)G_x+2\omega_x\sigma_0^2 G_r}{(\sigma_0^2+\omega_0^2-\omega_x^2)^2+(2\sigma_0 \omega_x)^2}.
    \end{eqnarray}
    
    The oscillatory resonance occurs at zero-crossing frequencies of $\lambda_{zi,x}$, i.e., $\lambda_{zi,x}(\omega_x)=0$, \begin{eqnarray}
    \omega_{x1,x2}=\dfrac{2G_r \sigma_0 \pm \sqrt{(2G_r\sigma_0)^2+ 4(\sigma_0^2+\omega_0^2) G_x^2}}{2 G_x},  
    \label{eq:T_wr}
    \end{eqnarray} \noindent where the feasible solutions correspond to positive zero-crossing frequency values with the largest magnitude~\cite{Liu2018AnRenewables}.

    In case of poorly damped oscillatory modes where $\abs{\sigma_0}<<\abs{\omega_0}$, it implies that $\omega_x$ approximately matches with the frequency of the oscillatory mode $\omega_0$, i.e., $\omega_x \approx \omega_0$. Thus, the real part of $\lambda_{zi}$ at $\omega_x$ can be approximated as \begin{eqnarray}
    \lambda_{z0r}(\omega_{x}\approx \omega_0) \approx \dfrac{-2 G_x \omega_0 \sigma_0}{2 \omega_0 \sigma_0} =\dfrac{-G_x}{(2 \omega_0 \sigma_0)^2}= k_x \sigma_0,
    \label{eq:T_larw}
    \end{eqnarray}
    
    \noindent where $k_x$ is the slope of $\lambda_{z0x}$ at $\omega_x \approx \omega_0$, i.e., \begin{eqnarray}
    k_x=\left[ \dfrac{\partial \lambda_{zix}(\omega)}{\partial \omega} \right]_{\omega=\omega_x} \approx
    \dfrac{-8 \omega_x^3 G_x \sigma_0^2-8 \omega_x^2 G_r \sigma_0^3}{16 \omega_0^4 \sigma_0^4} \approx \dfrac{- G_x}{2 \omega_0 \sigma_0^2}.
    \end{eqnarray}
    
    According to the above, the PMD stability criterion is summarized as follows,
    
    \vspace{0.2cm}
    
    \noindent \emph{PMD stability criterion}: multi-infeed grid-connected VSCs systems are stable (i.e., $\sigma_0<0$) if and only if,\\ \vspace{-0.5cm}\begin{center}(i) $k_x>0$ and $\lambda_{zi,r}<0$; or (ii) $k_x<0$ and $\lambda_{zi,r}>0$ \\ \end{center} at resonant frequencies $\omega_x$ for all local maximums or peak values of $\abs{\lambda_{zi}(j\omega)}$ ($i$~=~1~to~$n$). 
    
    \vspace{0.2cm}
    
     The condition $k_x > 0$ indicates that $\lambda_{zi,x}(\omega)$ passes through zero-axis at $\omega_x$ from a capacitive area to an inductive area; and the condition $k_x < 0$ means that $\lambda_{zi,x}(\omega)$ passes through zero-axis at $\omega_x$ from an inductive area to a capacitive area. 
     
     The second condition usually occurs at peak resonance points for inductive (i.e., positive $\lambda_{zi,x}(\omega)$ values which increase in line with the frequency, $j\omega L$) and capacitive (i.e., negative $\lambda_{zi,x}(\omega)$ values which decrease as long as the frequency increases, $-j/(\omega C)$) behaviour, which is associated to physical elements in conventional power systems. However, the control structure of power converter can also produce different inductive and capacitive behaviour (i.e., influence of the outer loops and the PLL) which is not related to any physical element of the system as observed in~\cite{Liu2017SubsynchronousNetworks} for PMSG based wind farms in weak AC networks in the subsynchronous frequency range. In this case, both conditions, (i) and (ii), might be considered for $\sigma_0<0$ at the peak resonance points.
     
    In the harmonic range as studied in Section~\ref{sec:Multi-infeed}, the imaginary part of the VSC output impedance is not strongly affected by the control structure and keeps the inductive behaviour produced by its filter, $L_c$. In this case, the condition (ii) might happen for $\sigma_0<0$ at the peak resonance points, which is the usual case in traditional electrical power systems. 
    
    
\section{Case study}

\begin{figure}[htbp]
    \centering
    \subfigure[]{\includegraphics[width=0.45\linewidth]{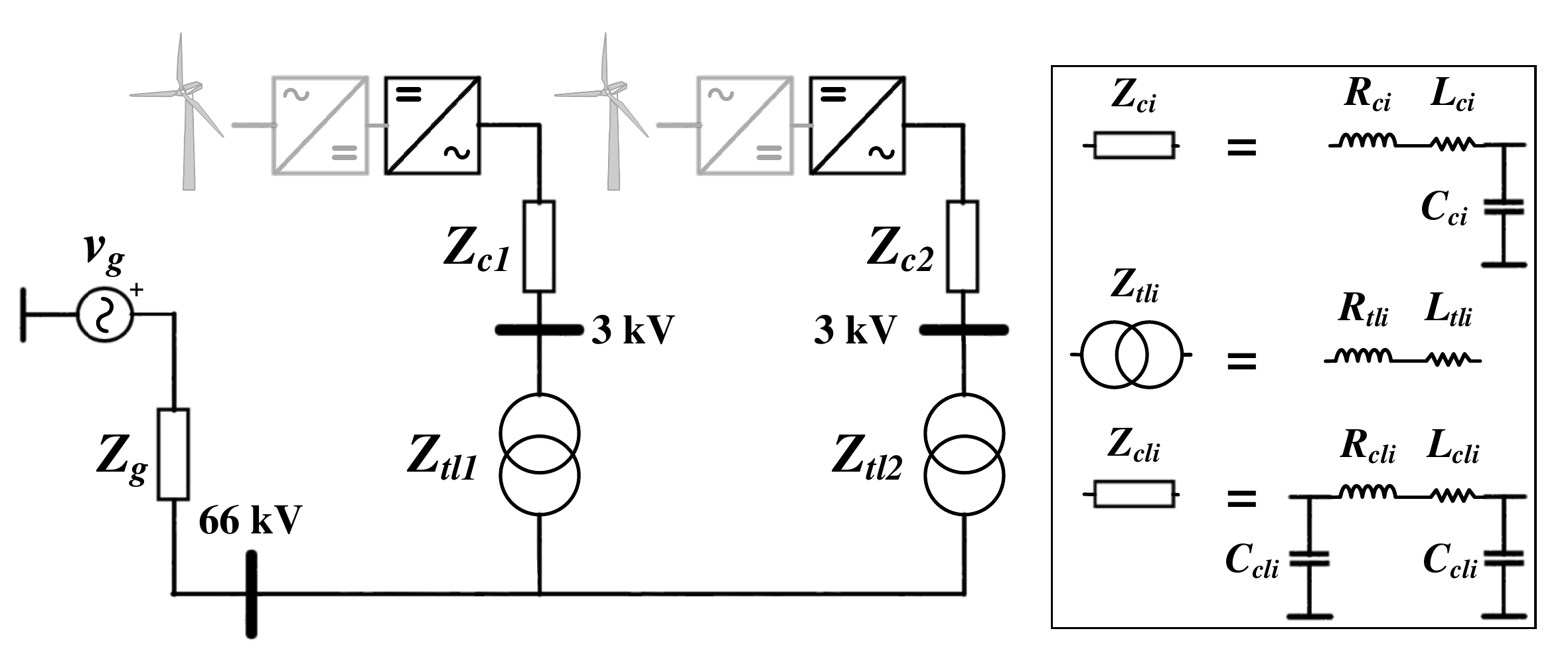}\label{subfig:T_tcsi}}
    \subfigure[]{\includegraphics[width=0.45\linewidth]{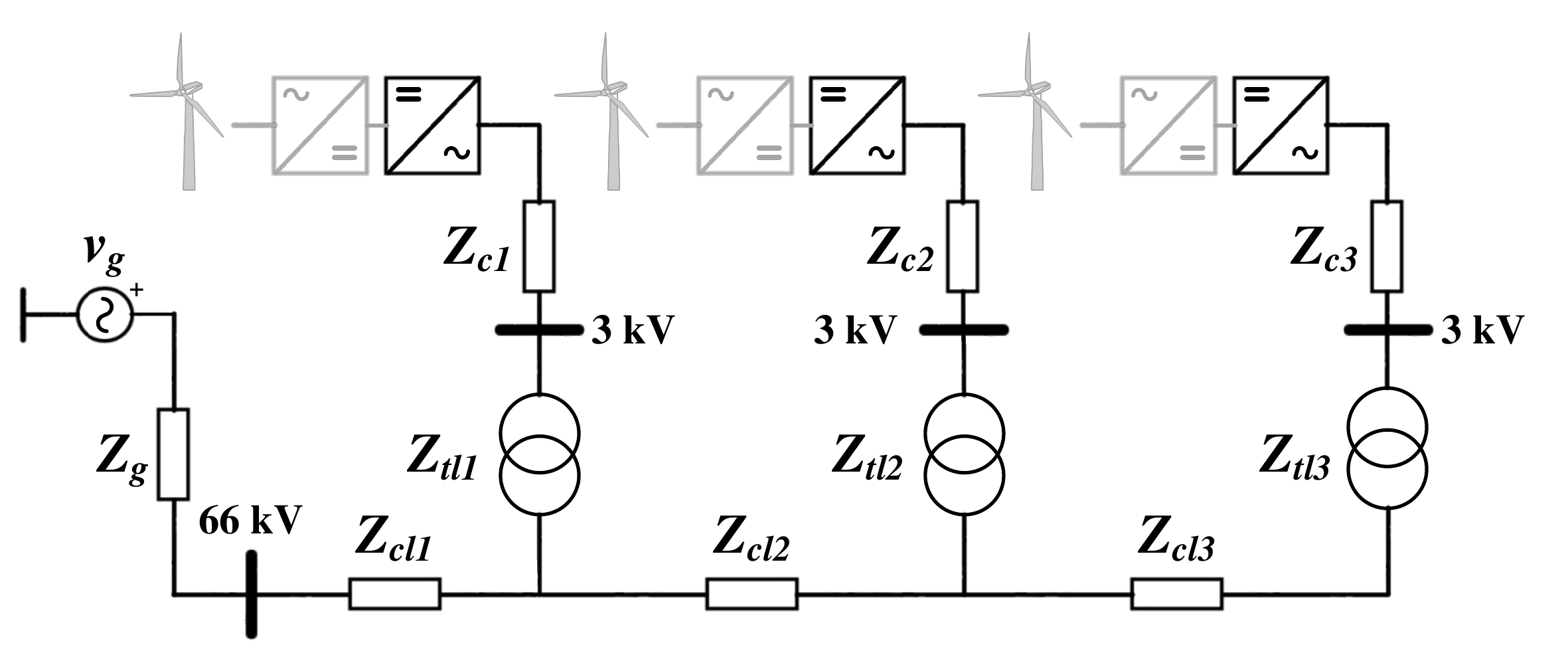}\label{subfig:T_tcsii}}
    \subfigure[]{\includegraphics[width=0.5\linewidth]{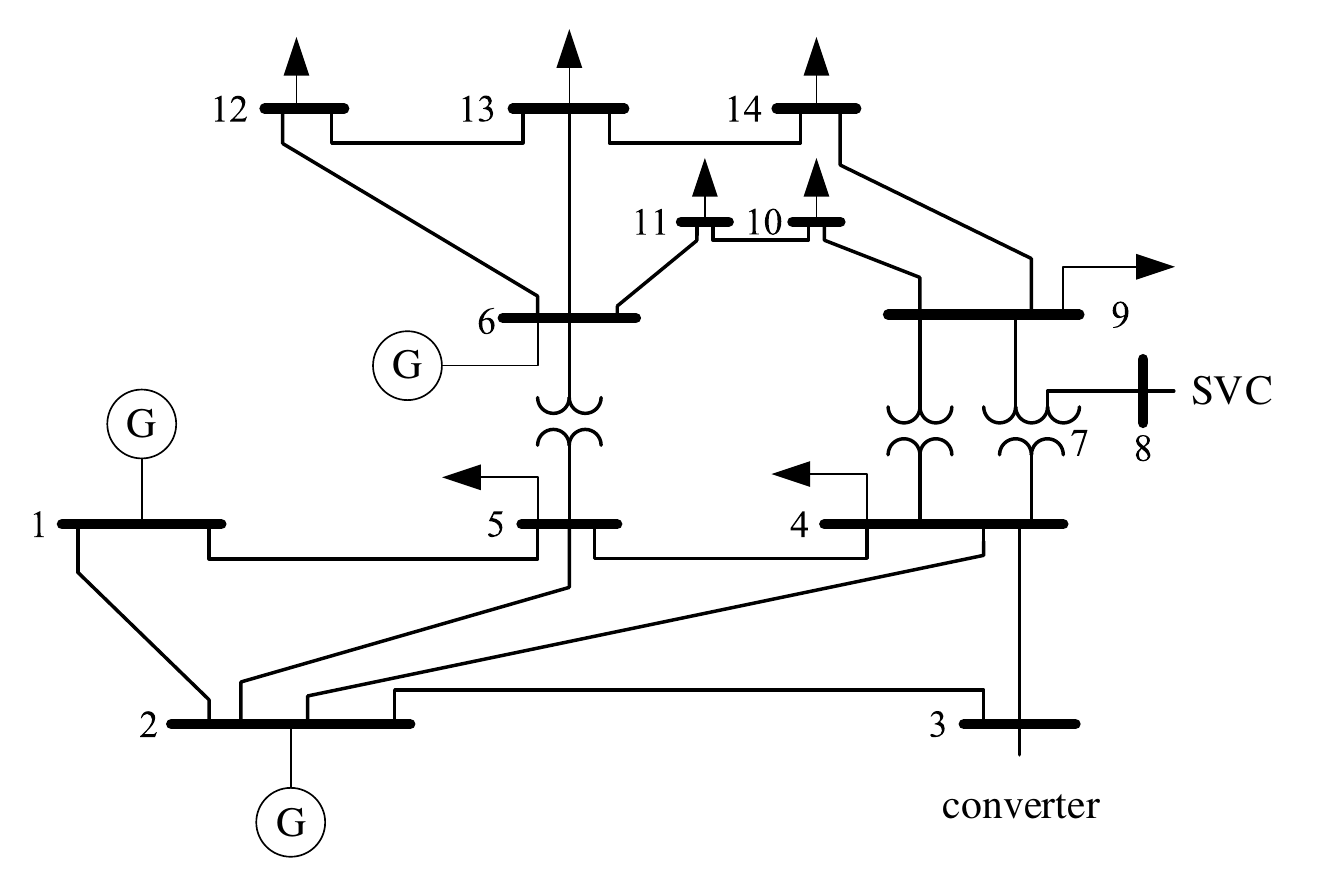}\label{subfig:T_tcsiii}}
    \caption{Testing networks single-line diagrams. (a) Case study I. (b) Case study II(a). (c) Case study II(b).}
    \label{fig:T_tcs}
    \end{figure}

    The previously described stability criteria is tested in two study cases: 
    \begin{itemize}[leftmargin=*]
         
    \item Case study I studies the effect of misleading association when dividing the system to assess the closed-loop stability. The study network consists of two VSCs connected in parallel to a  grid-equivalent impedance as shown Fig.~\ref{subfig:T_tcsi}. 
    
    \item Case study II addresses the issue of assessing the stability of a large power system in the frequency domain by studying two networks. 
    
    (a) The testing network of case study I is taken a step ahead by completing the string configuration with a 2~km cable between converters. An additional VSC is also connected in string as displayed in Fig.~\ref{subfig:T_tcsii}. 
    
    (b) A larger and more complex system than previous study networks as the modified IEEE 14 bus system (Fig.~\ref{subfig:T_tcsiii}) introduced in~\cite{Abu-hashim1999TestSimulation} is used to complete the study.
    
    \end{itemize}
    
    The construction of the nodal admittance matrix closed-loop of case study I and II(a) is derived in this work, and the dynamic models are verified in the $s$-domain by comparing the poles and zeros of the impedance matrix, $\mathbf{Z_T}(s)=(\mathbf{Y_T}(s))^{-1}$, with the eigenvalues of linear state-space models. The state-space models are validated with time domain simulations by comparing the results obtained with the ones of non-linear Simulink models. No state-space model was developed for case study II(b) because the process can be long and complex. Nevertheless, this example allows to verify the usefulness of the proposed stability criterion in large networks and compare its performance to the GNC.
    
    The stability assessment with the PMD stability criterion is compared with the results of eingenvalue analysis and the GNC in linearized state-space and impedance-based Matlab models for case study I and II(a), and only in an impedance-based model for case study II(b). The unstable resonance modes oscillation frequency of linear models is further verified for all study cases with time domain simulations of non-linear Simulink models. The system (complemented with data from~\cite{PROMOTioN2016DeliverableWPPs}) and control parameters for both networks can be found in Table~\ref{tab:Parameters}.  
    
    \begin{table}[htbp]
    \centering
    \caption{System parameters}\renewcommand\arraystretch{1}
    \label{tab:CA-PARAM}
    \vspace{-0.3cm}
    \begin{tabular}[c]{lccccl}
    \hline \hline
    Symbol       & Value & Units  & Symbol       & Value   & Units\\
    \hline
    $R_{c}$   & 0.0112   & $\Omega$ & $kp_{pll}$   & 0.0163    & rad / V s\\
    $L_{c}$   & 0.358    & mH       & $ki_{pll}$   & 0.326     & rad / V $s^2$\\
    $C_{c}$   & 141.471  & $\mu$F   & $kp_{ol}$    & 4.0825e-6 & 1 / V\\ 
    $R_{tl}$  & 0.00557  & $\Omega$ & $ki_{ol}$    & 0.00408   & 1 / V s\\
    $L_{tl}$  & 0.184    & mH       & $kp_{il}$    & 0.358     & H / s\\
    $R_{cl}$  & 9.773e-4 & $\Omega$ & $ki_{il}$    & 11.25     & $\Omega$ / s\\
    $L_{cl}$  & 0.00182  & mH       & $\tau_{ffv}$ & 0.010     & s\\
    $C_{cl}$  & 82.28    & $\mu$F   & $\tau_{fd}$  & 1.250e-4  & s\\
    \hline \hline
    \label{tab:Parameters}
    \end{tabular}
    \vspace{-0.3cm}
    \end{table}
 
    The time delay is calculated with the expression $\tau_{fd}=q_{d} \tau_{sw}$ as described in~\cite{Harnefors2007ModelingMatrices}. A $q_{d}$= 0.25 was initially considered for all converters in all study cases. The switching period $\tau_{sw}=1/f_{sw}$ is determined for $f_{sw}=2$ kHz. Instability happens when the time delay is modified (i.e., varying $q_d$)~\cite{Sainz2019AdmittanceInstabilities}.

    \subsection{Case study I}
    
\begin{figure*}[htbp]
    \subfigure[]{\includegraphics[width=0.33\linewidth]{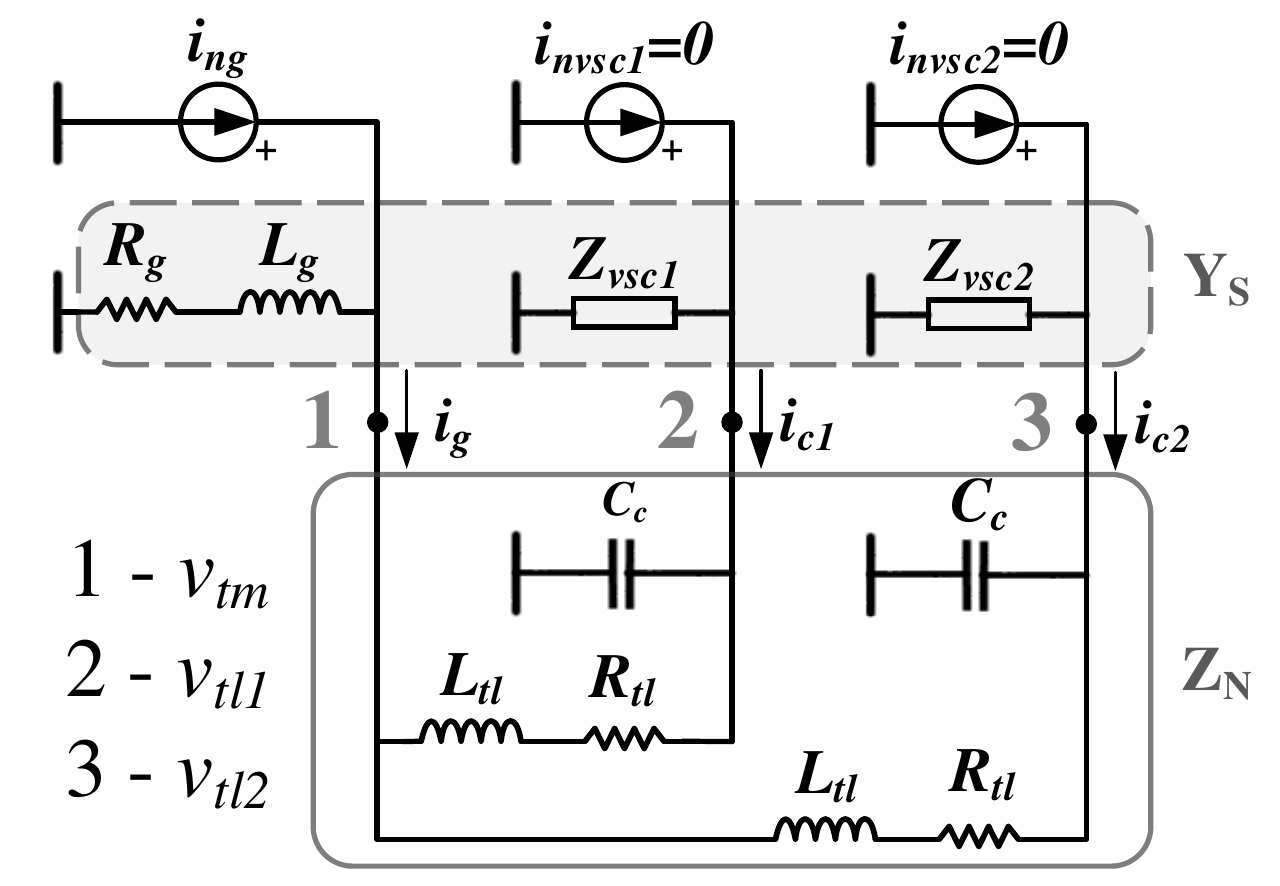}\label{subfig:T_tni}}
    \subfigure[]{\includegraphics[width=0.33\linewidth]{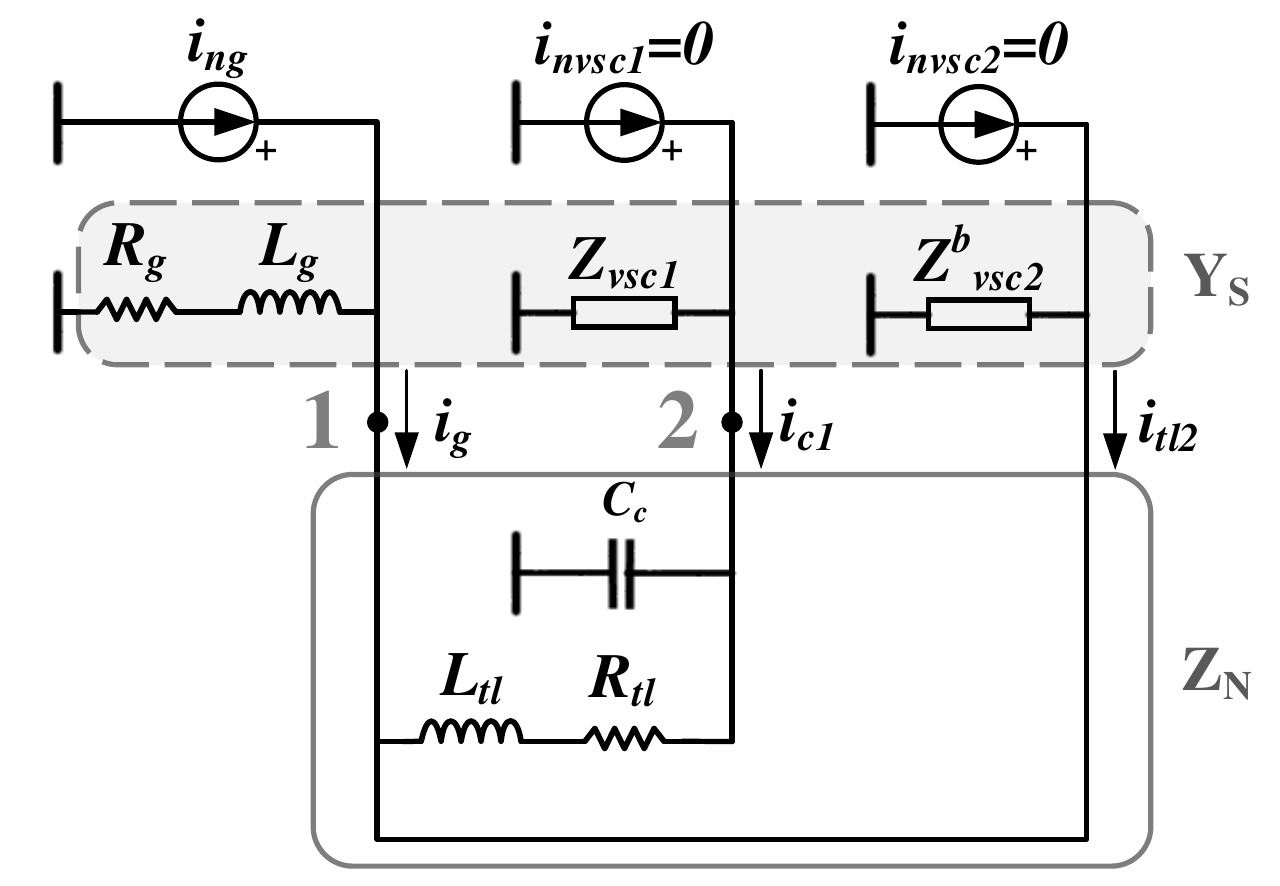}\label{subfig:T_tnii}}
    \subfigure[]{\includegraphics[width=0.33\linewidth]{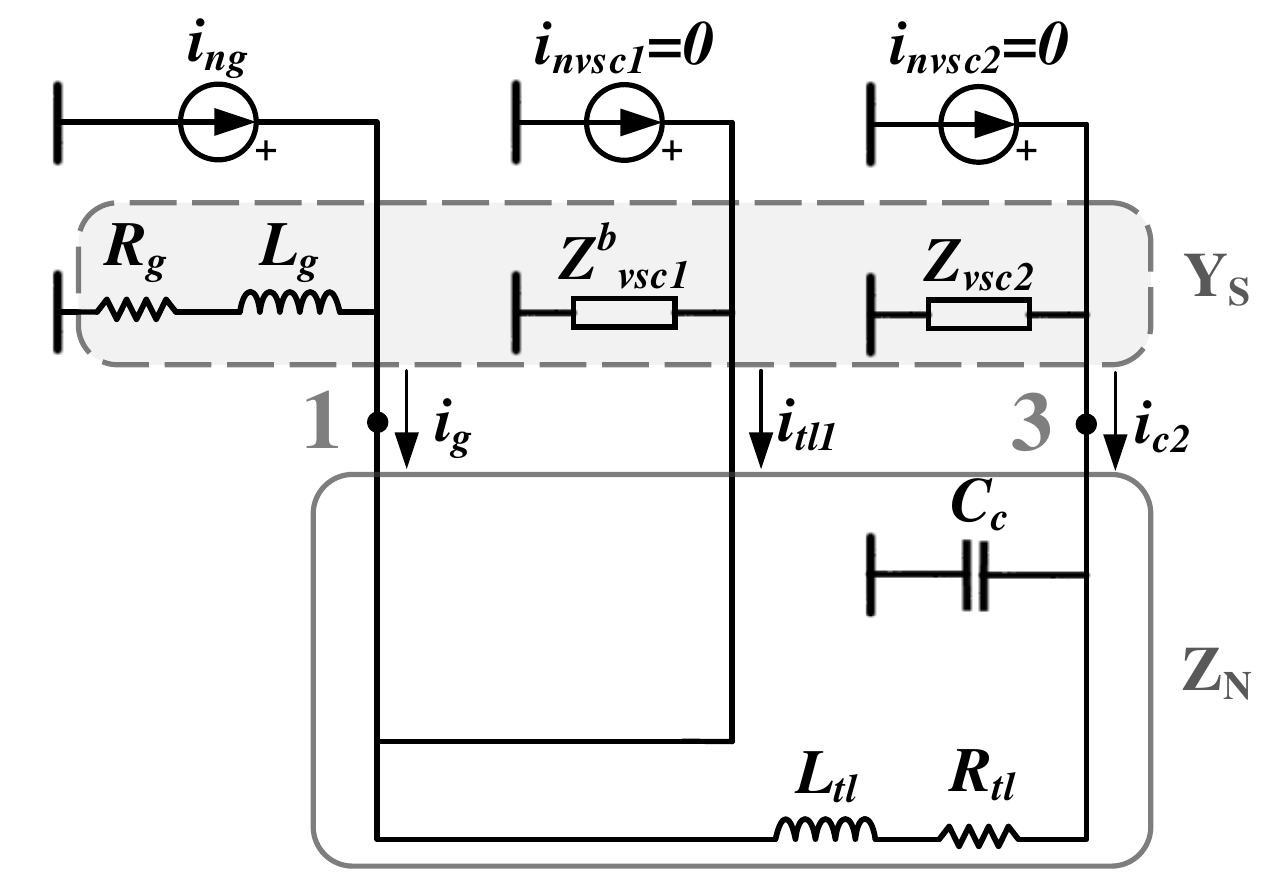}\label{subfig:T_tniii}}
    \caption{Case study I, grouping options. (a) GO1. (b) GO2. (c) GO3.} 
    \label{fig:T_tn}
    \end{figure*}

    The testing network is a 3 bus system with 2 converters (i.e., VSC1 and  and VSC2) connected to a network equivalent impedance (Fig~\ref{subfig:T_tcsi}). The system stability is assessed for three grouping options. Each of them associates some elements in $\mathbf{Y_S}$ and $\mathbf{Z_N}$ of the closed-loop function with a different approach as displayed in  Fig~\ref{fig:T_tn}. The nodal admittance matrix model has been verified in the time domain and the $s$-domain, where there is a good match with non-linear Simulink and linearized state-space models for stable (i.e., figures are not included for the sake of space) and unstable conditions in Fig~\ref{subfig:T_tdi} and Fig~\ref{subfig:T_pzi}. Instability in the system occurs in the time domain simulation at approximately 1190 Hz when the time delay of VSC2 increases up to 0.5 times $\tau_{sw}$ as displayed Fig~\ref{subfig:T_tdi}.

    \subsubsection{Grouping option 1} The nodal admittance matrix in the grouping option 1 (GO1) is constructed according to Section~\ref{sec:Multi-infeed} and displayed in Fig.~\ref{subfig:T_tni}. The grid equivalent is connected at bus 1, and two VSCs at buses 2 and 3 respectively. The network passive elements are grouped in $\mathbf{Y_N}$ and external elements which can cause instability by interacting with resonances of the system in $\mathbf{Y_S}$ (i.e., no open-loop RHP poles). The system closed-loop as expressed in~\eqref{eq:T_ztii} is conformed by \begin{eqnarray}
    \mathbf{Z_N} =
    \begin{bmatrix}
    \textbf{Y}_{tl1}+\textbf{Y}_{tl2}  & -\textbf{Y}_{tl1}                  & -\textbf{Y}_{tl2}\\
    -\textbf{Y}_{tl1}  & \textbf{Y}_{tl1}+\textbf{Y}_{cc1}  & \textbf{0}_{2\times2}   \\
    -\textbf{Y}_{tl2}  & \textbf{0}_{2\times2}                    & \textbf{Y}_{tl2}+\textbf{Y}_{cc2} \\
    \end{bmatrix}^{-1}
    \end{eqnarray}
    
    \noindent and \begin{eqnarray}
    \mathbf{Y_S} = 
    \begin{bmatrix}
    \textbf{Y}_g     & \textbf{0}_{2\times2}   & \textbf{0}_{2\times2}\\
    \textbf{0}_{2\times2} & \textbf{Y}_{vsc1}  & \textbf{0}_{2\times2}\\
    \textbf{0}_{2\times2} & \textbf{0}_{2\times2}   &\textbf{Y}_{vsc2} \\
    \end{bmatrix},
    \end{eqnarray}
    
    \noindent where $\mathbf{Y_{tl1}}$, $\mathbf{Y_{tl2}}$, $\mathbf{Y_{cc1}}$, $\mathbf{Y_{cc2}}$, $\mathbf{Y_{vsc1}}$, $\mathbf{Y_{vsc2}}$ and $\mathbf{Y_{g}}$ are 2 by 2 matrices, which can be expressed in the frequency domain or $s$-domain as in~\eqref{eq:T_z}. The inputs and outputs of the system are $\Delta i_n=[\Delta i_{ng-q} \enspace \Delta i_{ng-d} \enspace 0_{2\times1} \enspace 0_{2\times1}]^T$ and $\Delta v=[\Delta v_{tmv-q}$ \enspace $\Delta v_{tmv-d}$ \enspace $\Delta v_{tl1-q}$ \enspace $\Delta v_{tl1-d}$ \enspace $\Delta v_{tl2-q}$ \enspace $\Delta v_{tl2-d}]^T$ respectively.
    
\begin{figure*}[htbp]
    \centering
    \subfigure[]{\includegraphics[width=0.259\linewidth]{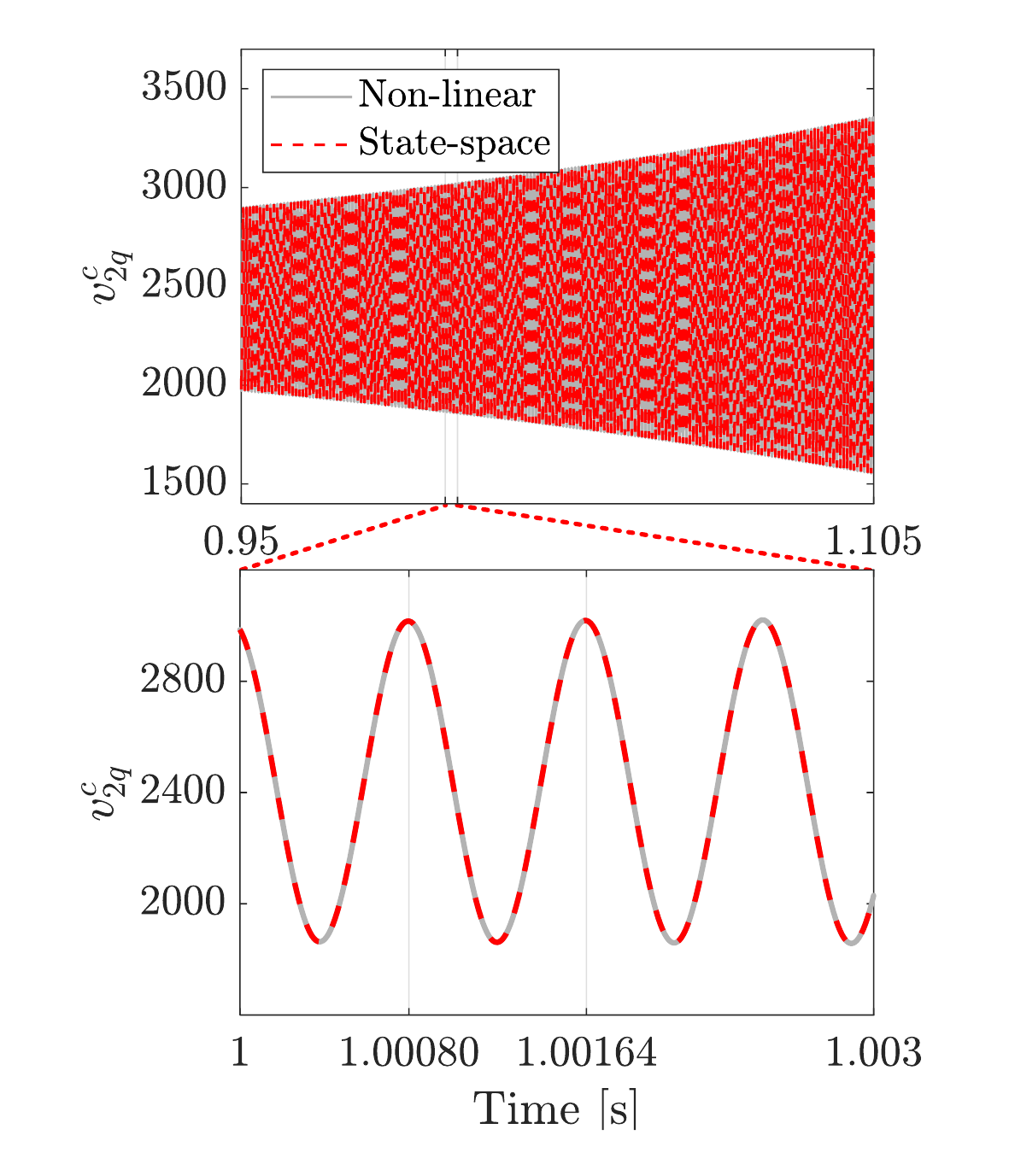}\label{subfig:T_tdi}}
    \subfigure[]{\includegraphics[width=0.259\linewidth]{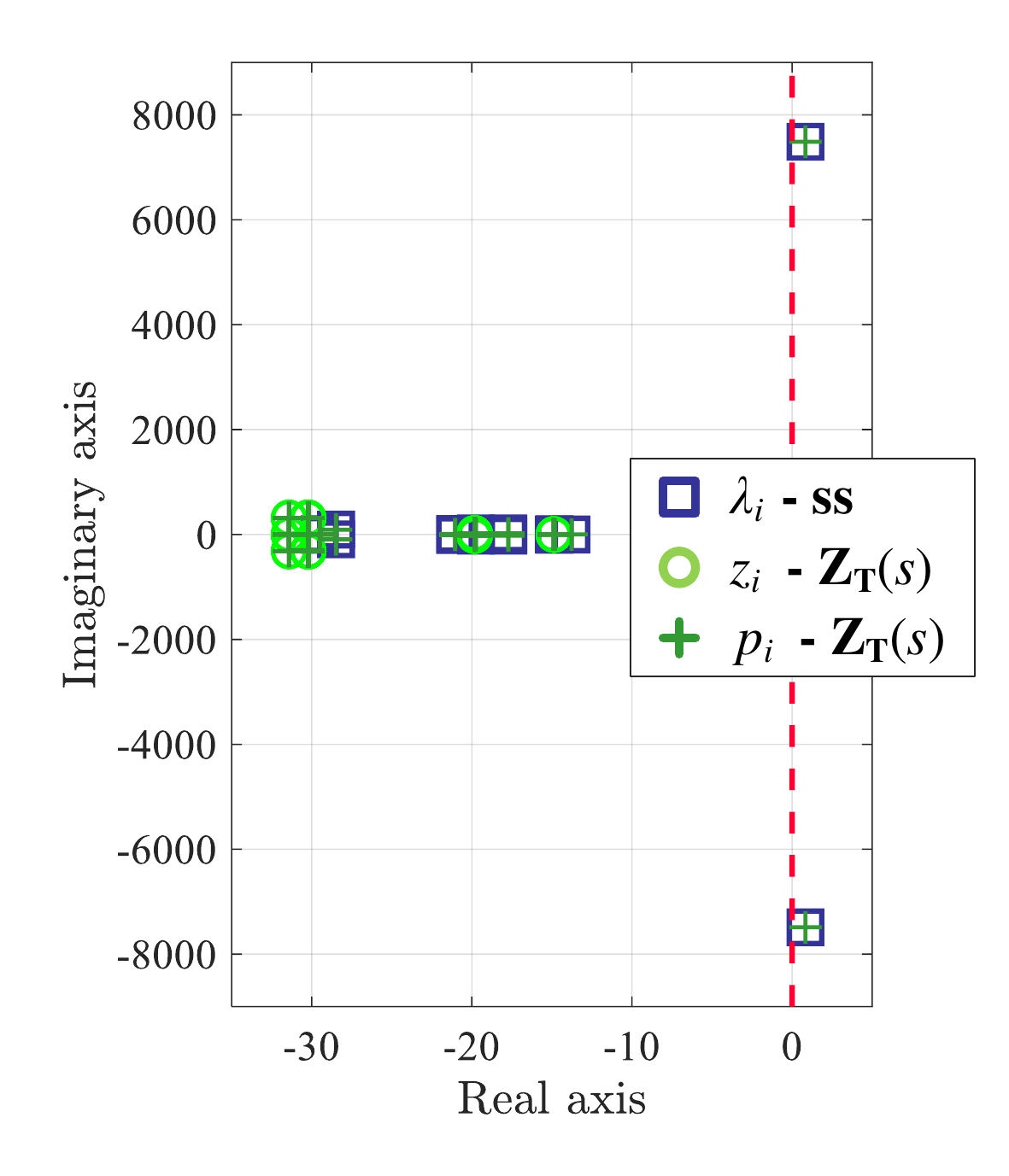}\label{subfig:T_pzi}}
    \subfigure[]{\includegraphics[width=0.469\linewidth]{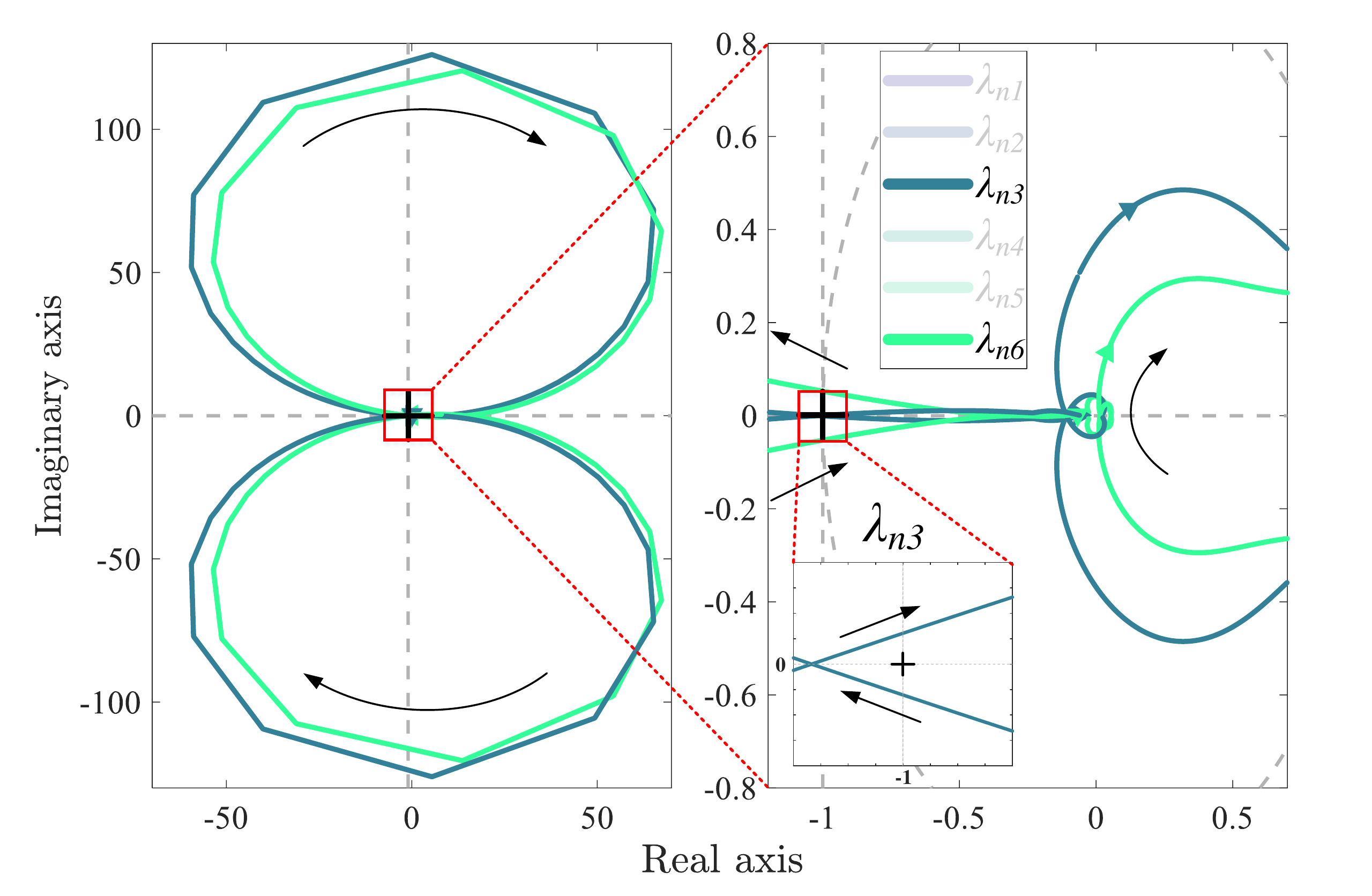}\label{subfig:T_nyqi}}
    \caption{Stability assessment GO1. (a) Time domain simulation. (b) Eigenvalue analysis. (c) GNC.} 
    \label{fig:T_ssassia}
    \end{figure*}

    In Fig.~\ref{fig:T_ssassia}, the stability is assessed with $s$-domain and frequency domain stability criteria. A pair complex conjugate poles in the RHP can be noticed in Fig.~\ref{subfig:T_pzi} at $f_0$ = 1192 Hz (i.e., $\omega_0 = 7488 = 2 \pi f_0$) by evaluating the system impedance matrix $\mathbf{Z_T}(s)$. These poles match with the eigenvalues of the state-space representation of $\mathbf{Z_T}$ as described in the eigenvalue analysis section. The instability can be further confirmed in Fig.~\ref{subfig:T_nyqi}, where the $\lambda_{n3}$ Nyquist curve of $\mathbf{L}$ encircles the (-1, $j$0) point in the clockwise direction. It is worth mentioning that other Nyquist curves such as $\lambda_{n6}$ seem to encircle (-1, $j$0) but by zooming around the critical point no encirclement was observed.
    
\begin{figure}[htbp]
    \centering
    {\includegraphics[width=0.5\linewidth]{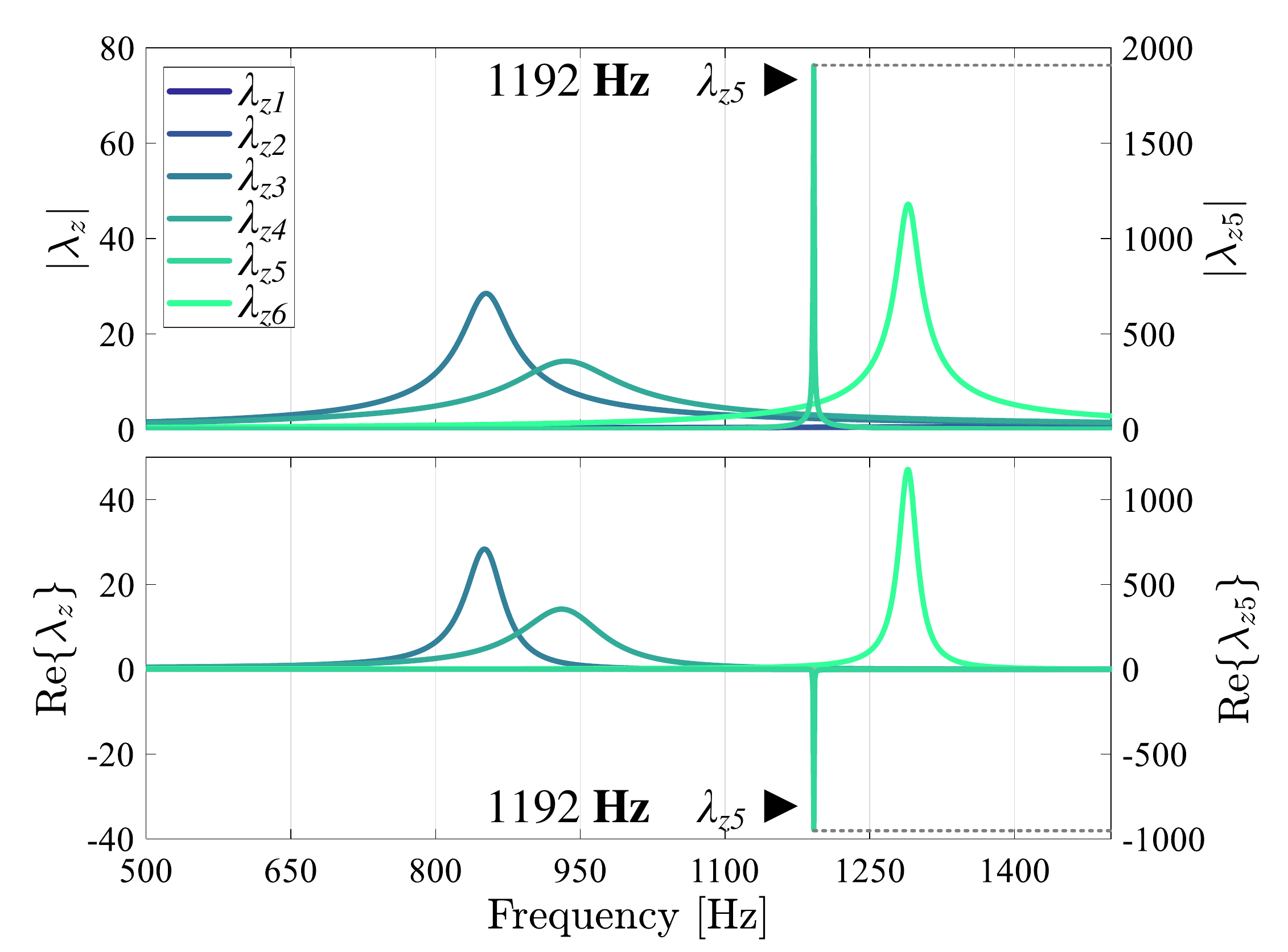}\label{subfig:T_fdi}}
    \caption{Stability assessment GO1, PMD stability criterion.} 
    \label{fig:T_ssassib}
    \end{figure}

    The stability assessment with the PMD stability criterion is displayed in Fig.~\ref{fig:T_ssassib}. The modal impedance magnitude curve \textbf{$\lambda_{z5}$} in the frequency domain has a peak at 1192 [Hz] where its real part is negative, confirming once more the instability of $\mathbf{Z_T}$. The stability assessment in GO1 agrees for all stability criteria. 
    
    \subsubsection{Grouping option 2} The grouping option 2 (GO2) studies the same network under the same instability conditions as GO1. However, it merges $\mathbf{Y}_{cc2}$ and $\mathbf{Y}_{tl2}$ into $\mathbf{Y}_{vsc2}$ as displayed in Fig.~\ref{subfig:T_tnii}. The new converter admittance is $\mathbf{Y}_{vsc2}^{b}=(\mathbf{Z}_{vsc2}^{b})^{-1}=[(\mathbf{Z}_{vsc2}//\mathbf{Z}_{cc2})+\mathbf{Z}_{tl2}]^{-1}$ which enables the possibility of interaction between the converter controller with resonant circuits from the network (e.g., parallel resonances caused by the transformer inductance $\mathbf{Z}_{tl2}$ and the shunt capacitor of the converter filter $\mathbf{Z}_{cc2}$)  within $\mathbf{Y}_{vsc2}^{b}$. In consequence, the system matrix order is reduced (i.e., the number of buses of the network is reduced from three to two), and the closed-loop is conformed by \begin{eqnarray}
    \mathbf{Z_N}= 
    \begin{bmatrix}
    \textbf{Y}_{tl1} & -\textbf{Y}_{tl1} \\
    -\textbf{Y}_{tl1}  & \textbf{Y}_{tl1}+\textbf{Y}_{cc1}  \\
    \end{bmatrix}^{-1}
    \end{eqnarray}
    
    \noindent and  \begin{eqnarray}
    \mathbf{Y_S}=   
    \begin{bmatrix}
    \textbf{Y}_g + \textbf{Y}_{vsc2}^b    & \textbf{0}_{2\times2}\\
    \textbf{0}_{2\times2} & \textbf{Y}_{vsc1}\\
    \end{bmatrix},
    \end{eqnarray}
    
    \noindent where the inputs and outputs of the system are $\Delta i_n=[\Delta i_{ng-q} \enspace \Delta i_{ng-d} \enspace 0_{2\times1}]^T$ and $\Delta v=[\Delta v_{tmv-q}$ \enspace $\Delta v_{tmv-d}$ \enspace $\Delta v_{tl1-q}$ \enspace $\Delta v_{tl1-d}]^T$.
    
\begin{figure}[htbp]
    \subfigure[]{\includegraphics[width=0.5
    \linewidth]{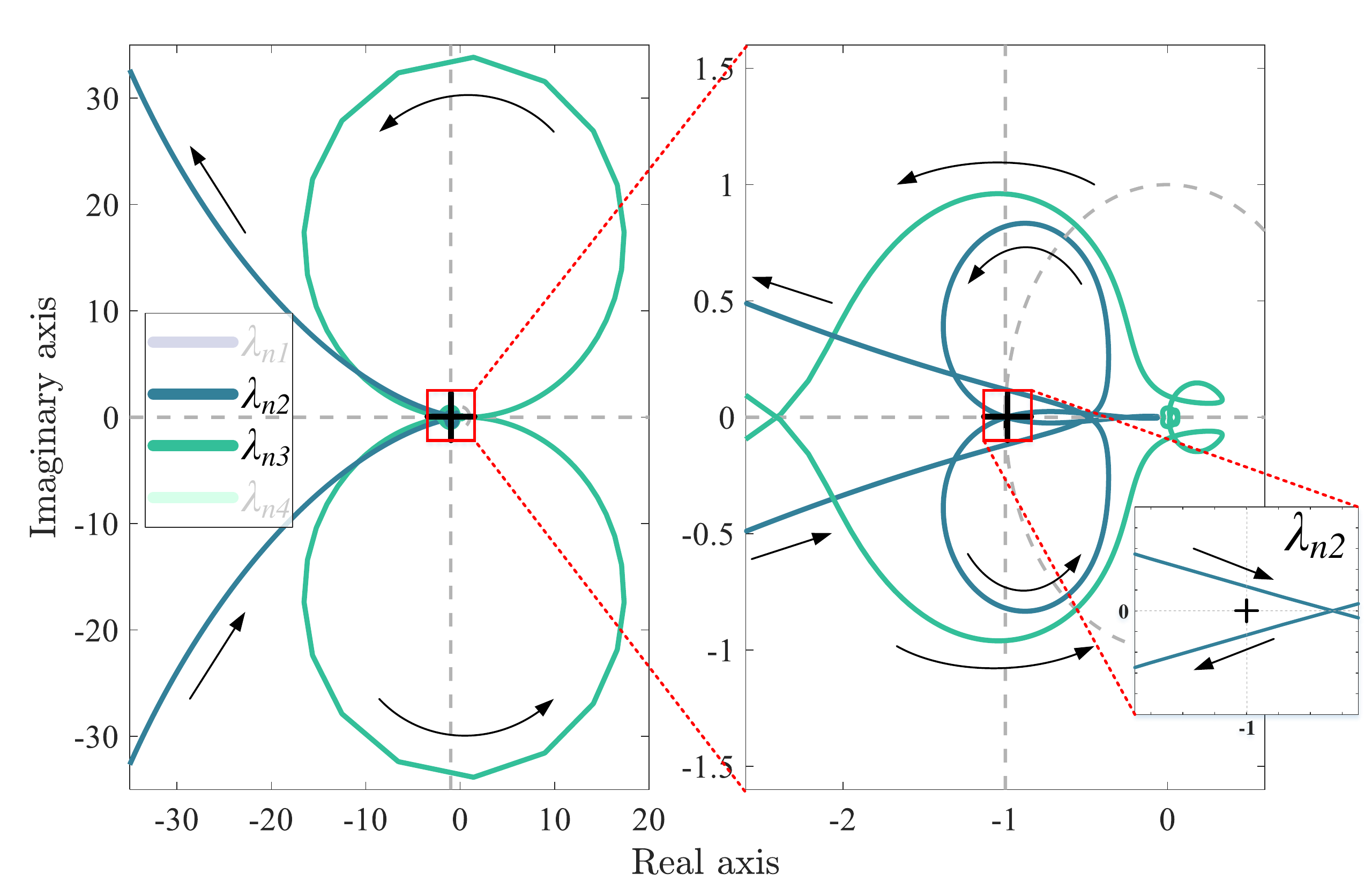}\label{subfig:T_nyqii}}
    \subfigure[]{\includegraphics[width=0.5\linewidth]{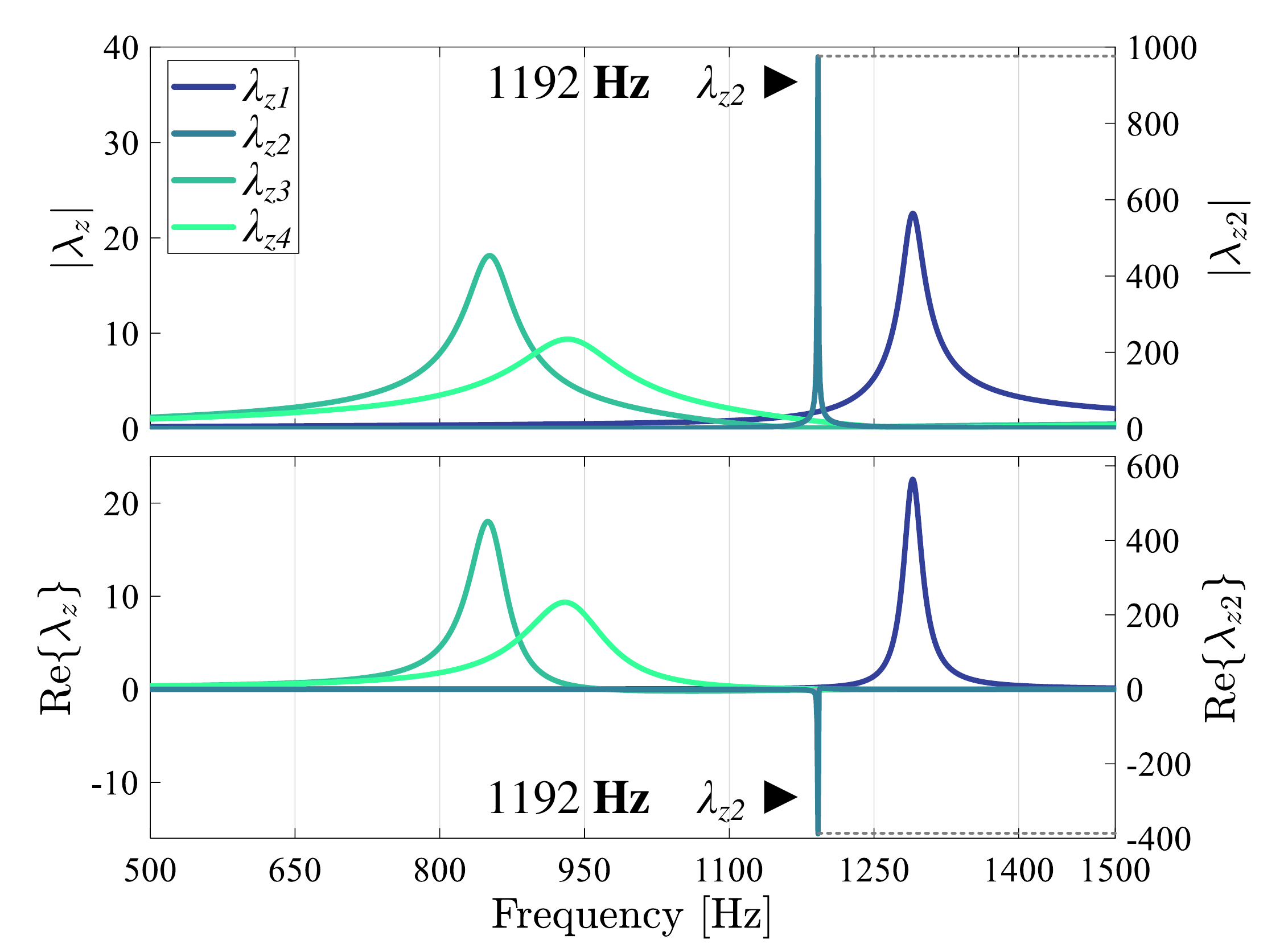}\label{subfig:T_fdii}}
    \caption{Stability assessment GO2. (a) GNC. (b) PMD stability criterion.} 
    \label{fig:T_ssassiia}
    \end{figure}

    The stability of GO2 is assessed with the GNC in Fig.~\ref{subfig:T_nyqii}. The $\lambda_{n3}$ Nyquist curve encircles the (-1, $j$0) point two times but in counterclockwise direction. On the other hand, a closed-up view around the critical point shows that $\lambda_{n2}$ is not enclosing it. In GO1, the system was identified as unstable for all three criteria, but the GNC criterion fails to predict stability in this grouping option. This is caused by the  open-loop RHP poles introduced by $\mathbf{Y_{vsc2}^b}$ in $\mathbf{Y_S}$.
    
    The PMD stability criterion is tested in GO2 as displayed in Fig.~\ref{subfig:T_fdii}. There is a peak in one of the magnitude of the eigenvalues curves $\abs{ \lambda_{z2} }$ at 1192 Hz where it matches with its negative real part, $\textrm{Re}{\{\lambda_{z2}\}}$, as it was obtained in GO1. The instability was once again confirmed by the PMD stability criterion, and the result was not affected by the RHP poles of $\mathbf{Y_S}(s)$.  The PMD stability criterion does not fail in the stability assessment of the GO2 because the stability criteria is applied to the closed-loop transfer function of the system.
    
    \subsubsection{Grouping option 3} The instability in the grouping option 3 (GO3) is caused by VSC2 as in previous grouping options, but the elements $\mathbf{Y}_{cc1}$ and $\mathbf{Y}_{tl1}$ are grouped into $\mathbf{Y}_{vsc1}$ as displayed in Fig.~\ref{subfig:T_tniii}. The order of the system matrix is also reduced, and the closed-loop is composed by \begin{eqnarray}
    \mathbf{Z_N} =
    \begin{bmatrix}
    \textbf{Y}_{tl2} & -\textbf{Y}_{tl2} \\
    -\textbf{Y}_{tl2}  & \textbf{Y}_{tl2}+\textbf{Y}_{cc2}  \\
    \end{bmatrix}^{-1}
    \end{eqnarray}
    
    \noindent and \begin{eqnarray}
    \mathbf{Y_S} = 
    \begin{bmatrix}
    \textbf{Y}_g + \textbf{Y}_{vsc1}^b     & \textbf{0}_{2\times2}\\
    \textbf{0}_{2\times2} & \textbf{Y}_{vsc2}\\
    \end{bmatrix},
    \end{eqnarray}
    
    \noindent where $\mathbf{Y}_{vsc1}^b=(\mathbf{Z}_{vsc1}^b)^{-1}=[(\mathbf{Z}_{vsc1}//\mathbf{Z}_{cc1})+\mathbf{Y}_{tl1}]^{-1}$, and  the inputs and outputs of the system are $\Delta i_n=[\Delta i_{ng-q} \enspace \Delta i_{ng-d} \enspace 0_{2x1}]^T$ and $\Delta v=[\Delta v_{tmv-q}$ \enspace $\Delta v_{tmv-d}$ \enspace $\Delta v_{tl2-q}$ \enspace $\Delta v_{tl2-d}]^T$.
    
\begin{figure}[htbp]
    \subfigure[]{\includegraphics[width=0.5\linewidth]{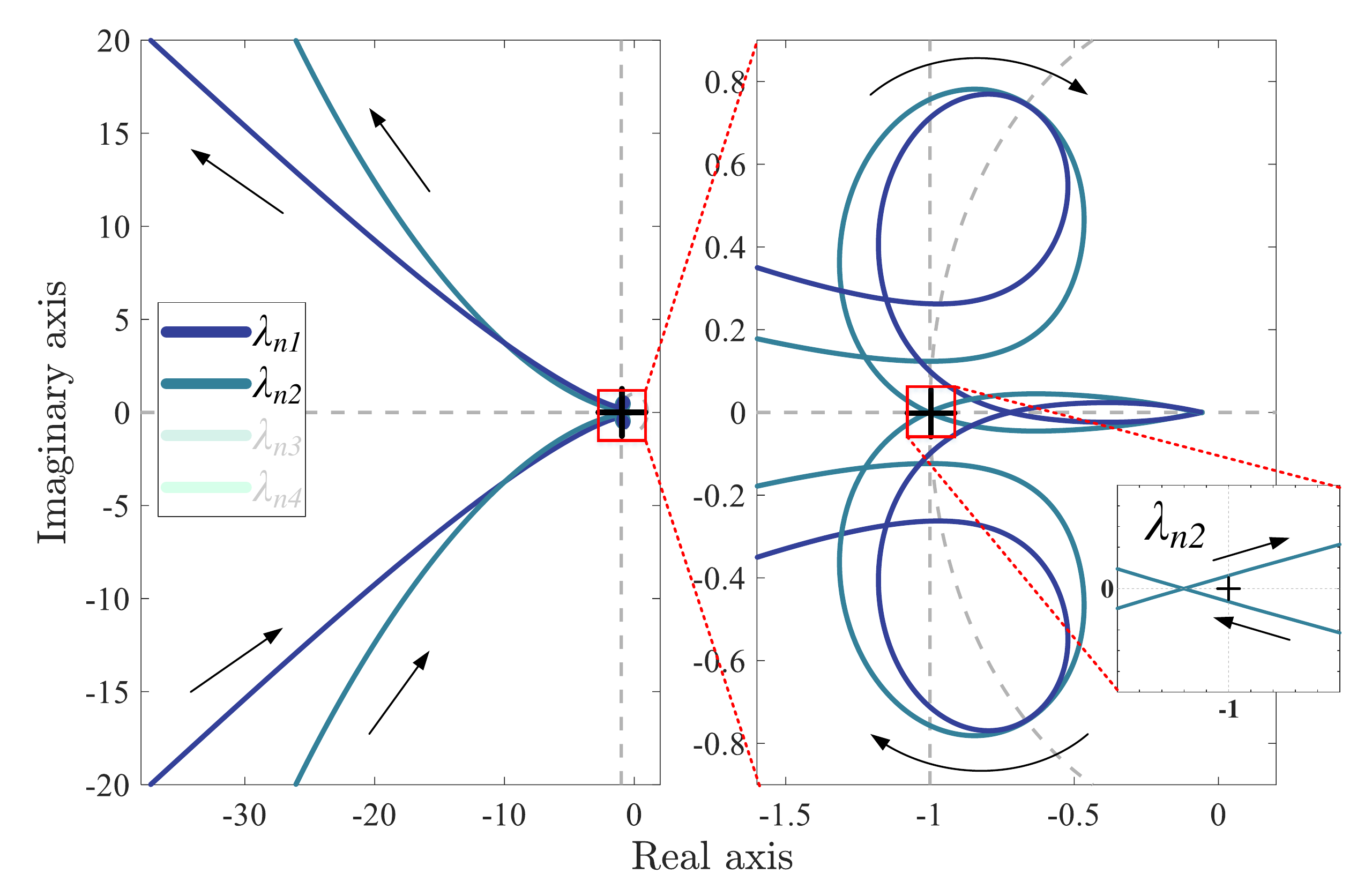}\label{subfig:T_nyqiii}}
    \subfigure[]{\includegraphics[width=0.5\linewidth]{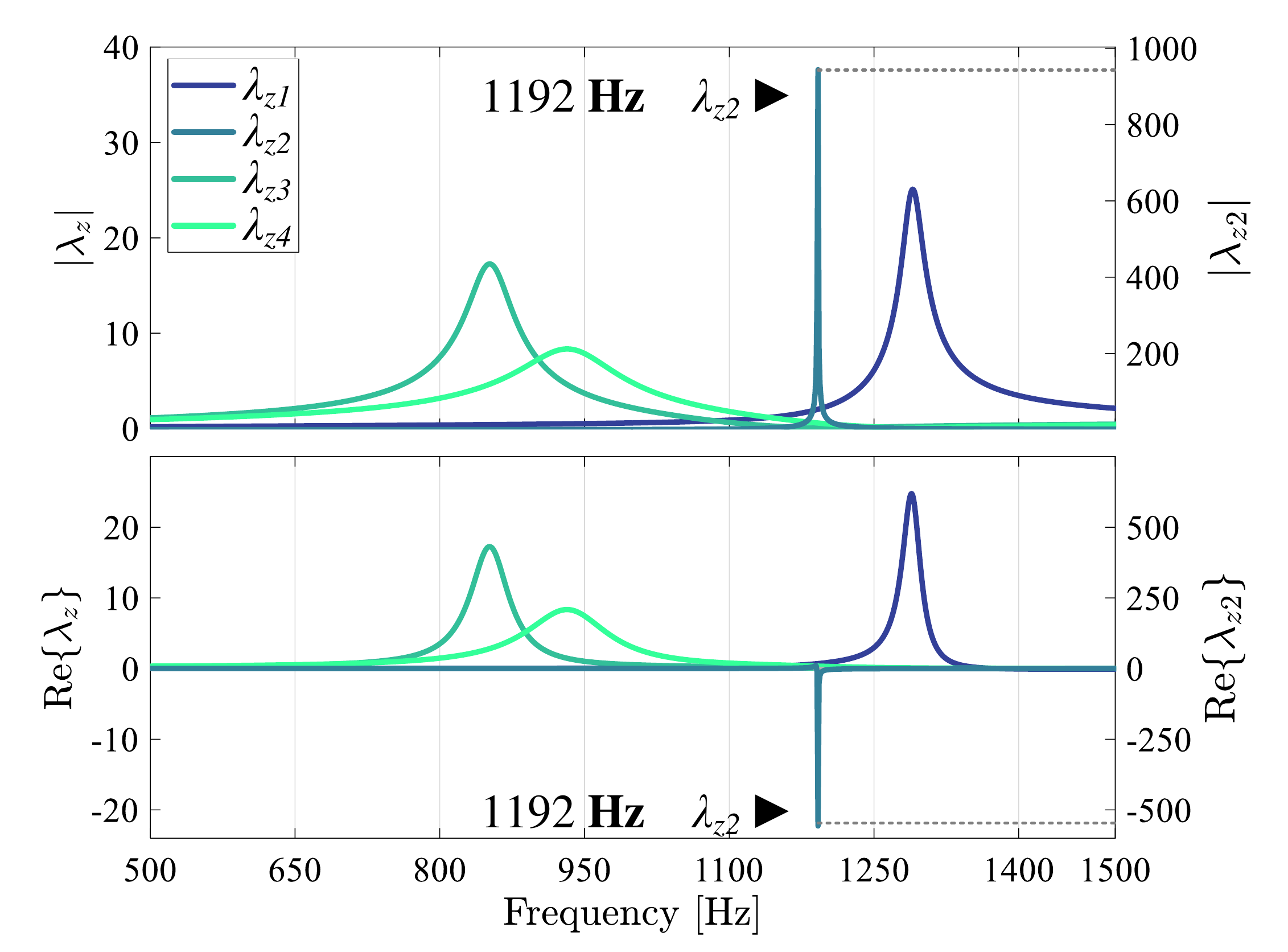}\label{subfig:T_fdiii}}
    \caption{Stability assessment GO3. (a) GNC. (b) PMD stability criterion.}
    \label{fig:T_ssassiiia}
    \end{figure}

    When addressing the stability of $\mathbf{Z_T}$ by evaluating the $\mathbf{L}$ with the GNC in Fig.~\ref{subfig:T_nyqiii}, the system instability was identified as GO1. The $\lambda_{n2}$ Nyquist curve encircles the critical point (-1, $j$0) in the clockwise direction. In GO3, no open-loop RHP poles were observed in comparison to GO2, because VSC2 is not merged with any grid component and the instability occurs due to the interaction between VSC2 and the grid.
    
    GO3 was also assessed with the PMD stability criterion in Fig.~\ref{subfig:T_fdiii}. The assessment matches the GNC one. Again, the instability was found at 1192 Hz as GO1 and GO2, where $\abs{\lambda_{z2}}$ in the frequency domain has a peak where its $\textrm{Re}{\{\lambda_{z2}\}}$ is negative.
    
    This case study proves the effectiveness of the proposed stability criterion PMD and the methodology in Section \ref{sec:Multi-infeed}. The malpractice at the moment of associating the elements of the network into two subsystems in order to study the closed-loop stability of $\mathbf{Z_T}$ could lead to misleading stability conclusions during the application of the GNC. 
    
    \subsection{Case study II}
    
     The effectiveness of the proposed method can be further tested in two larger networks than the one from case study I, example (a) and (b), where the GNC becomes more challenging due to the number of eigenvalue Nyquist curves.
     
    \subsubsection{Example (a)}
    
    The network of case study I is extended to add more complexity to the study. This network is composed by three VSCs connected in string configuration to a grid equivalent, as displayed in Fig.~\ref{subfig:T_tcsii}.
    
\begin{figure*}[htbp]
    \subfigure[]{\includegraphics[width=0.4\linewidth]{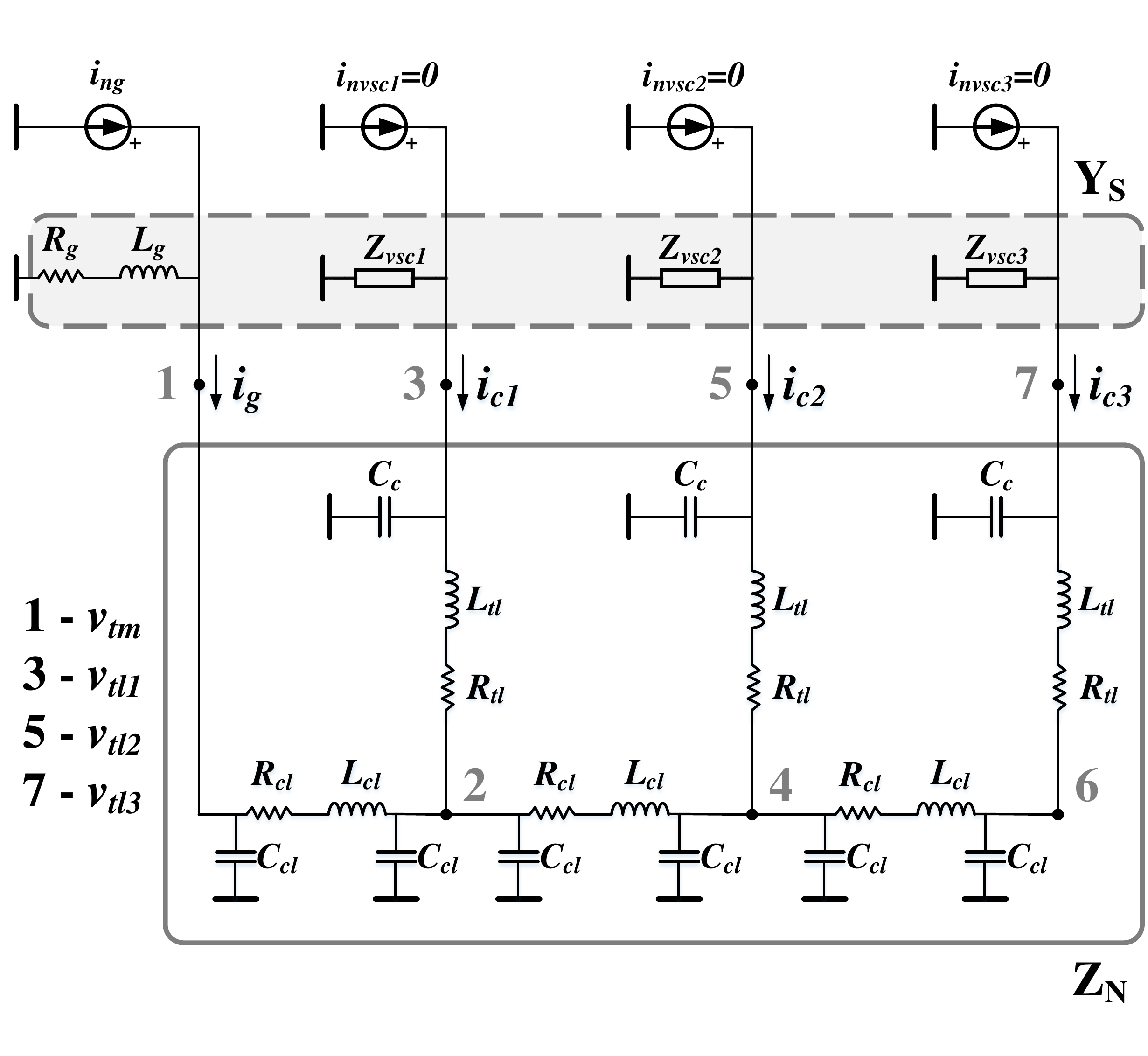}\label{subfig:T_tmi}}
    \subfigure[]{\includegraphics[width=0.3\linewidth]{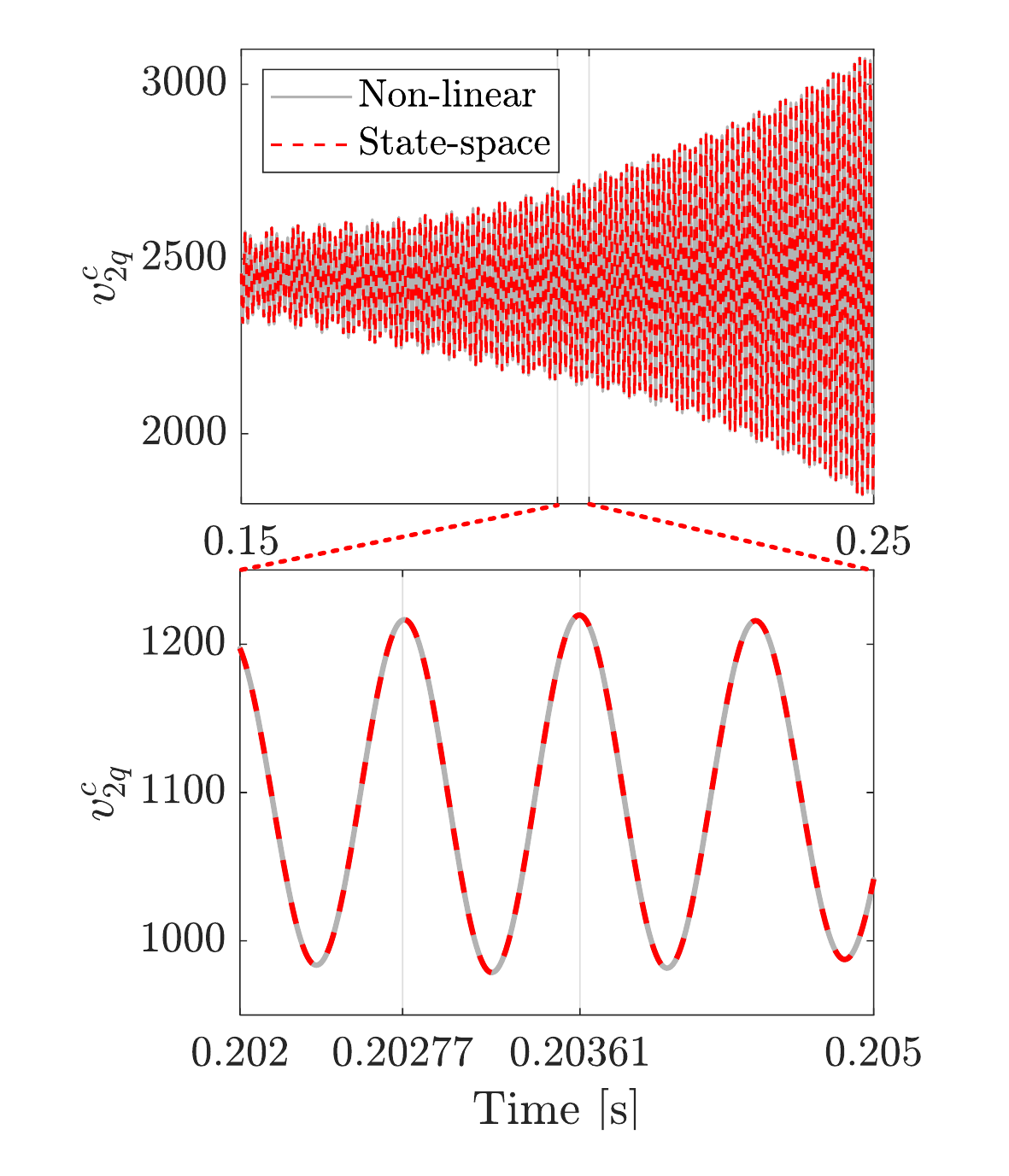}\label{subfig:T_tdii}}
    \subfigure[]{\includegraphics[width=0.3\linewidth]{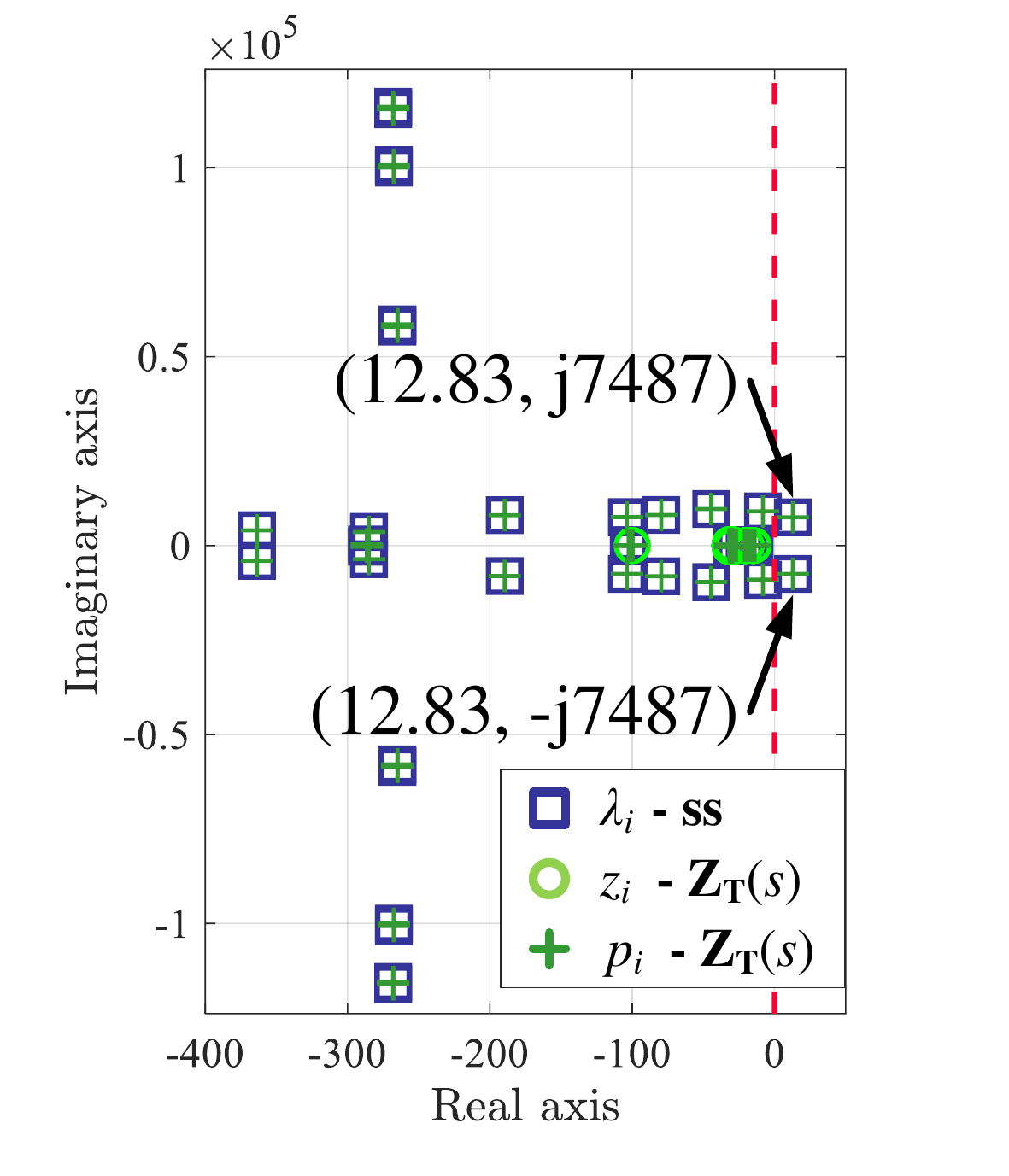}\label{subfig:T_pziv}}
    \subfigure[]{\includegraphics[width=0.5\linewidth]{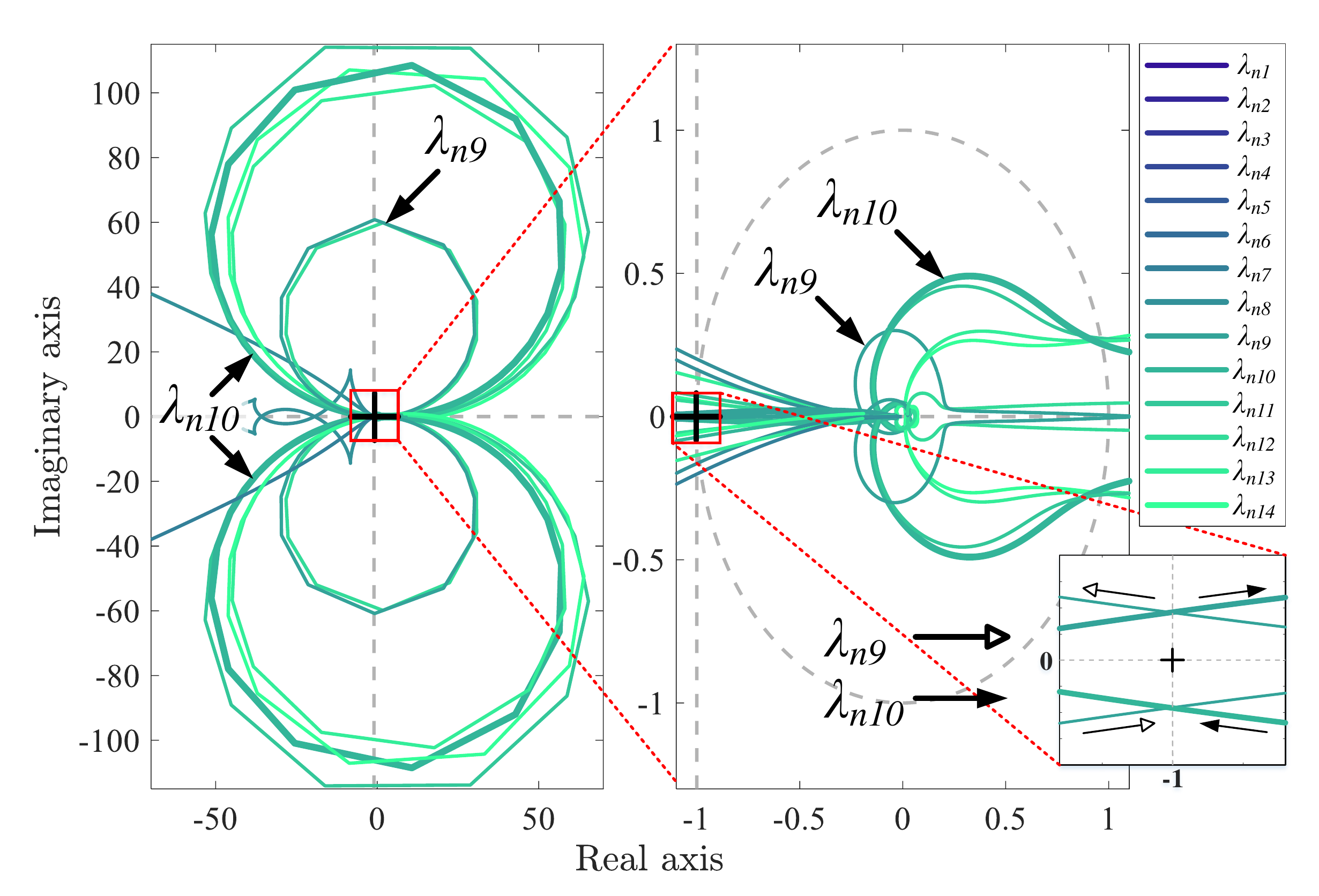}\label{subfig:T_nyqiv}}
    \subfigure[]{\includegraphics[width=0.5\linewidth]{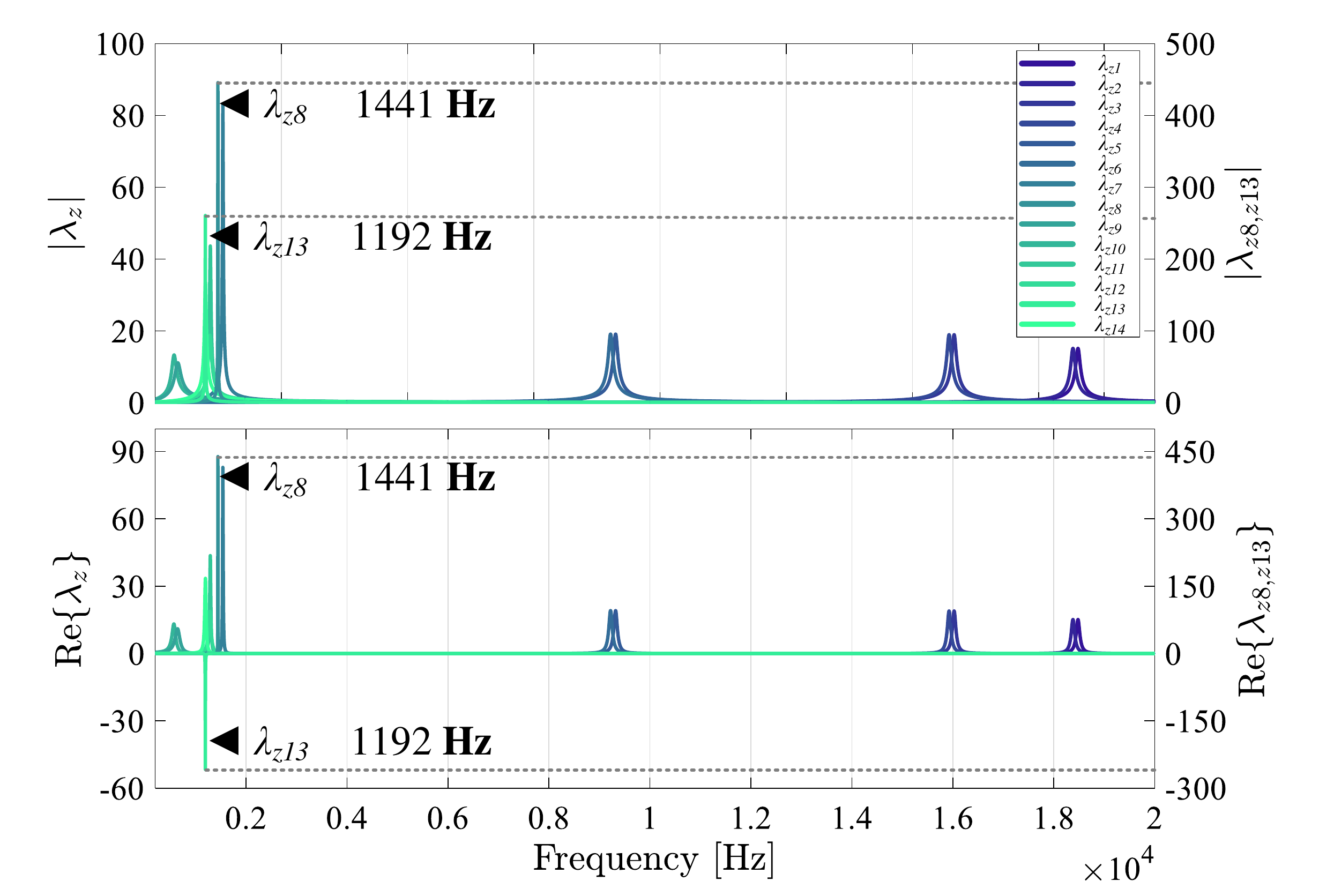}\label{subfig:T_fdiv}}
    \caption{Stability assessment case study II(a). (a) $\mathbf{Z_N}$ \& $\mathbf{Y_S}$. (b) Time domain simulation. (c) Eigenvalue analysis. (d) GNC. (e) PMD stability criterion.} 
    \label{fig:T_ssassiva}
    \end{figure*}

    The system is a 7 bus network where the grid equivalent is connected at bus 1, and the three converters are connected at buses 3, 5 and 7 respectively as illustrated in Fig.~\ref{subfig:T_tmi}. Each converter is separated from another by a 2 km cable modelled with a PI section equivalent $\mathbf{Y}_{cli}$, which is composed by a RL section $\mathbf{Y}_{rli}$ and a shunt capacitor $\mathbf{Y}_{ccli}$ on both ends. The elements of the network are associated as stated in Section~\ref{sec:Multi-infeed} to construct the closed-loop representation of $\mathbf{Z_T}$. The matrices $\mathbf{Y_S}$ and $\mathbf{Z_N}$ are detailed in~\eqref{eq:T_ziv} and~\eqref{eq:T_yiv}, where $\mathbf{Y}_{rl1-2}=\mathbf{Y}_{rl1}+\mathbf{Y}_{rl2}$; $\mathbf{Y}_{ccl1-2}=\mathbf{Y}_{ccl1}+\mathbf{Y}_{ccl2}$; $\mathbf{Y}_{rl2-3}=\mathbf{Y}_{rl2}+\mathbf{Y}_{rl3}$; $\mathbf{Y}_{ccl2-3}=\mathbf{Y}_{ccl2}+\mathbf{Y}_{ccl3}$; and the inputs and outputs are $\Delta i_n=[\Delta i_{ng-q} \enspace \Delta i_{ng-d} \enspace 0_{1x2} \enspace 0_{1x2} \enspace 0_{1x2}]^T$ and $\Delta v=[\Delta v_{tmv-q}$ \enspace $\Delta v_{tmv-d}$ \enspace $\Delta v_{tl1-q}$ \enspace $\Delta v_{tl1-d}$ \enspace $\Delta v_{tl2-q}$ \enspace $\Delta v_{tl2-d}$ \enspace $\Delta v_{tl3-q}$ \enspace $\Delta v_{tl3-d}]^T$.
    
    \begin{eqnarray}
    \mathbf{Y_S} =
    \begin{bmatrix}
    \mathbf{Y}_g & \mathbf{0}_{2\times2} & \mathbf{0}_{2\times2} & \mathbf{0}_{2\times2} & \mathbf{0}_{2\times2} & \mathbf{0}_{2\times2} & \mathbf{0}_{2\times2}\\     
    \mathbf{0}_{2\times2} & \mathbf{0}_{2\times2} & \mathbf{0}_{2\times2} & \mathbf{0}_{2\times2} & \mathbf{0}_{2\times2} & \mathbf{0}_{2\times2} & \mathbf{0}_{2\times2}\\   
    \mathbf{0}_{2\times2} & \mathbf{0}_{2\times2} & \mathbf{Y}_{vsc1} & \mathbf{0}_{2\times2} & \mathbf{0}_{2\times2} & \mathbf{0}_{2\times2} & \mathbf{0}_{2\times2}\\ 
    \mathbf{0}_{2\times2} & \mathbf{0}_{2\times2} & \mathbf{0}_{2\times2} & \mathbf{0}_{2\times2} & \mathbf{0}_{2\times2} & \mathbf{0}_{2\times2} & \mathbf{0}_{2\times2}\\  
    \mathbf{0}_{2\times2} & \mathbf{0}_{2\times2} & \mathbf{0}_{2\times2} & \mathbf{0}_{2\times2} & \mathbf{Y}_{vsc2} & \mathbf{0}_{2\times2} & \mathbf{0}_{2\times2}\\     
    \mathbf{0}_{2\times2} & \mathbf{0}_{2\times2} & \mathbf{0}_{2\times2} & \mathbf{0}_{2\times2} & \mathbf{0}_{2\times2} & \mathbf{0}_{2\times2} & \mathbf{0}_{2\times2}\\  
    \mathbf{0}_{2\times2} & \mathbf{0}_{2\times2} & \mathbf{0}_{2\times2} & \mathbf{0}_{2\times2} & \mathbf{0}_{2\times2} & \mathbf{0}_{2\times2} & \mathbf{Y}_{vsc3}\\
    \end{bmatrix}
    \label{eq:T_ziv}
    \end{eqnarray} 
    
    \begin{figure*}[htbp]
    \begin{eqnarray}
    \mathbf{Z_N} =
    \tiny{\begin{bmatrix}
    \mathbf{Y}_{rl1} + \mathbf{Y}_{ccl1} & -\mathbf{Y}_{rl1} & \mathbf{0}_{2\times2} & \mathbf{0}_{2\times2} & \mathbf{0}_{2\times2} & \mathbf{0}_{2\times2} & \mathbf{0}_{2\times2}\\
    -\mathbf{Y}_{rl1}  & \mathbf{Y}_{tl1} + \mathbf{Y}_{rl1-2} + \mathbf{Y}_{ccl1-2} & -\mathbf{Y}_{tl1}  & -\mathbf{Y}_{rl2} & \mathbf{0}_{2\times2} & -\mathbf{Y}_{tl1}   & \mathbf{0}_{2\times2}\\
    \mathbf{0}_{2\times2} & -\mathbf{Y}_{tl1}   & \mathbf{Y}_{tl1} + \mathbf{Y}_{cc1} &  \mathbf{0}_{2\times2} & \mathbf{0}_{2\times2} & \mathbf{0}_{2\times2} & \mathbf{0}_{2\times2}\\
    \mathbf{0}_{2\times2} & -\mathbf{Y}_{rl2} & \mathbf{0}_{2\times2} & \mathbf{Y}_{tl2} + \mathbf{Y}_{rl2-3} + \mathbf{Y}_{ccl2-3} & -\mathbf{Y}_{tl2} & -\mathbf{Y}_{rl3} & \mathbf{0}_{2\times2}\\
    \mathbf{0}_{2\times2} & \mathbf{0}_{2\times2}   & \mathbf{0}_{2\times2}  & -\mathbf{Y}_{tl2} & \mathbf{Y}_{tl2} + \mathbf{Y}_{cc2} & \mathbf{0}_{2\times2}   & \mathbf{0}_{2\times2}\\
    \mathbf{0}_{2\times2} & \mathbf{0}_{2\times2}  & \mathbf{0}_{2\times2}  & -\mathbf{Y}_{rl3} & \mathbf{0}_{2\times2}  & \mathbf{Y}_{tl3} + \mathbf{Y}_{rl3} + \mathbf{Y}_{ccl3} & -\mathbf{Y}_{tl3}\\
    \mathbf{0}_{2\times2} & \mathbf{0}_{2\times2}   & \mathbf{0}_{2\times2}  & \mathbf{0}_{2\times2} & \mathbf{0}_{2\times2} & -\mathbf{Y}_{tl2}   & \mathbf{Y}_{tl3} + \mathbf{Y}_{cc3}\\
    \end{bmatrix}}^{-1}
    \label{eq:T_yiv}
    \end{eqnarray} 
    \end{figure*}
    
    The dynamic model has been verified with time domain simulations and in the $s$-domain, where there is a good agreement with the results of  non-linear Simulink and linearized state-space models as displayed in Fig~\ref{subfig:T_tdii} and Fig~\ref{subfig:T_pziv}.
    In this example, the instability of the system is caused by VSC2 when the $\tau_{fd}$ is increased up to $q_{d}$ = 0.45 times $\tau_{sw}$. The stability of $\mathbf{Z_T}$ is assessed in Fig.~\ref{fig:T_ssassiva} with frequency and $s$-domain criteria. In Fig.~\ref{subfig:T_pziv}, $\mathbf{Z_T}(s)$ is evaluated in the $s$-domain, and  a pair complex conjugate poles in the RHP can be noticed at $f_0$ = 1192 Hz (i.e., $\omega_0 = 7487 = 2 \pi f_0$) which match the eigenvalues of the system state-space matrix. The instability is confirmed in the frequency domain in Fig.~\ref{subfig:T_nyqiv}, where the $\lambda_{n10}$ Nyquist curve of $\mathbf{L}$ encircles the critical point (-1, $j$0) in the clockwise direction. It is worth mentioning that assessing stability was challenging and time consuming due to the number of eigenvalues to be studied. For instance, $\lambda_{n9}$ encircles the critical point but in counterclockwise direction, and other eigenvalue Nyquist curves $\lambda_{n11}$ to $\lambda_{n14}$ follow a similar trajectory as $\lambda_{n10}$, but they do not encircle the critical point.
    
    The stability assessment with the PMD stability criterion is displayed in Fig.~\ref{subfig:T_fdiv}. The magnitude of $\lambda_{z13}$ curve in the frequency domain has a peak at 1192 [Hz] where its real part is negative, confirming $\mathbf{Z_T}$ instability. It can also be spotted a mode at 1441 Hz with larger magnitude than the unstable mode. however, its real part is positive which makes it stable. This larger magnitude means that the real part is close to zero and this eigenvalue may be a candidate to lead system to instability. 
    
    \subsubsection{Example (b)}
    
    The network size and complexity is further extended in this example. A commonly used network in the literature as the IEEE 14 modified bus system (Fig~\ref{subfig:T_tcsiii}) introduced in~\cite{Abu-hashim1999TestSimulation} is used. The network is modelled according to the reference, with the exception of filters, the converter, and the SVC. The filters are not modelled because they do not contribute significantly to the main resonance modes according to~\cite{Xu2005HarmonicAnalysis}. The converter and the SVC (Static Var Compensator), which are connected to bus 3 and 8, are modelled as VSCs and scaled according to the reference data. The system instability in the system happens when the VSC time delay increases up to 0.6 times the switching period of VSC1.
    
    There are two grouping options in this example. First, in GO1, the system nodal admittance matrix is constructed as described in Section III. Then, in GO2, some system elements are associated in order to cause RHP in the open loop, which is a condition for the GNC to fail. 
    
    \paragraph{Grouping option 1}
    
\begin{figure*}[htbp]
    \centering
 
    \subfigure[]{\includegraphics[width=0.68\linewidth]{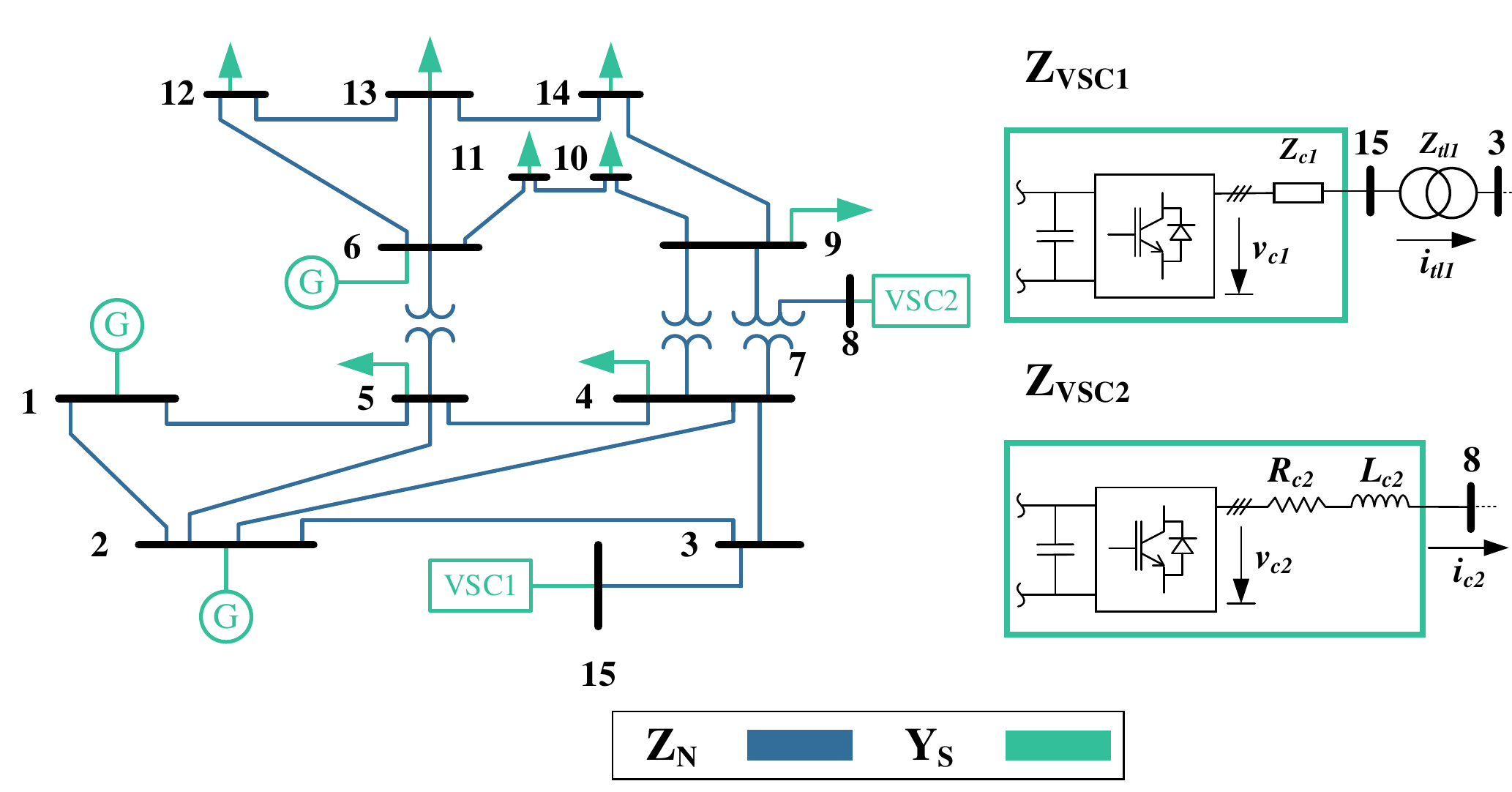}\label{subfig:T_toib}}
    \subfigure[]{\includegraphics[width=0.3\linewidth]{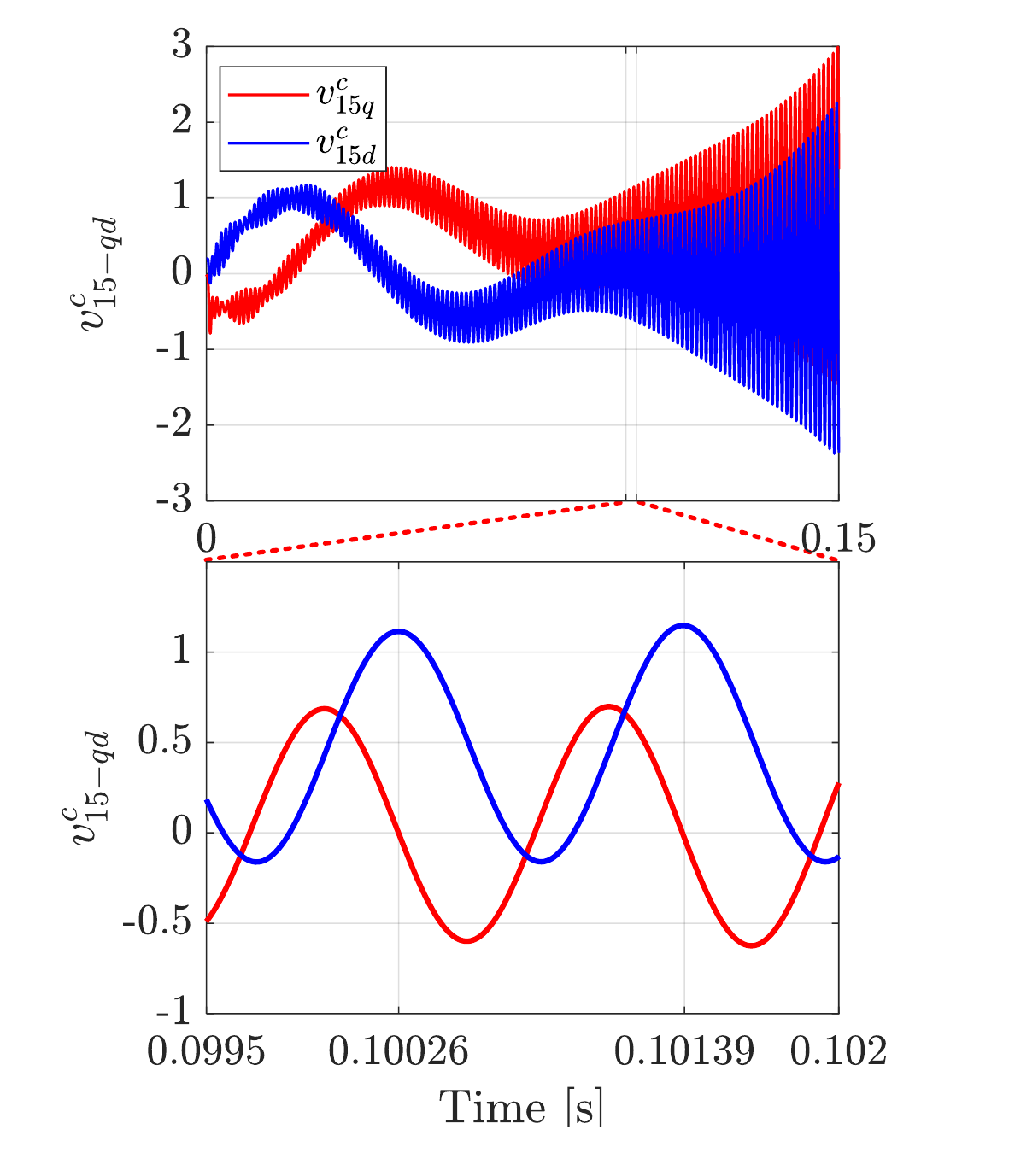}\label{subfig:T_tdv}}
    \subfigure[]{\includegraphics[width=0.42\linewidth]{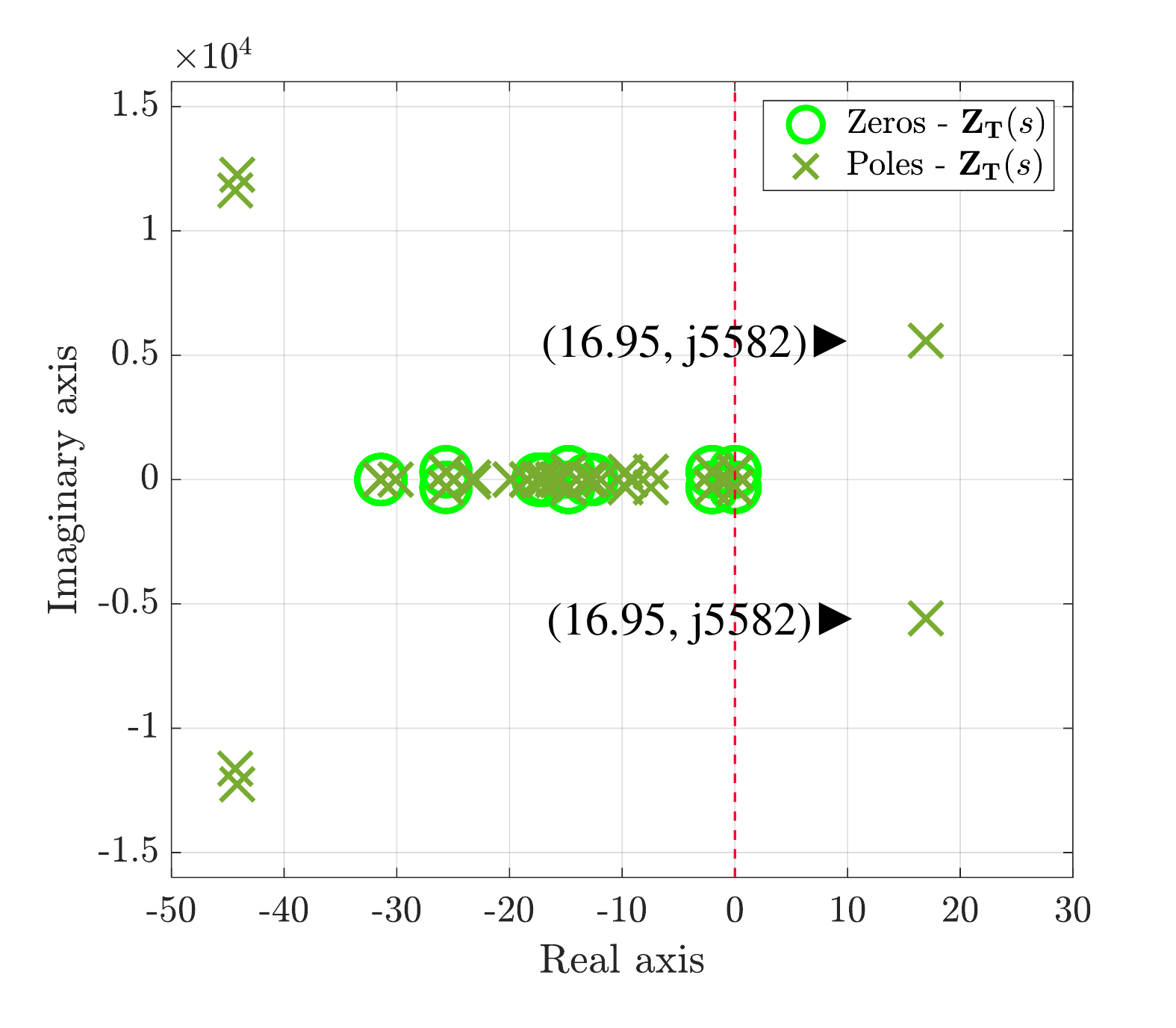}\label{subfig:T_pzv}}
    \subfigure[]{\includegraphics[width=0.5\linewidth]{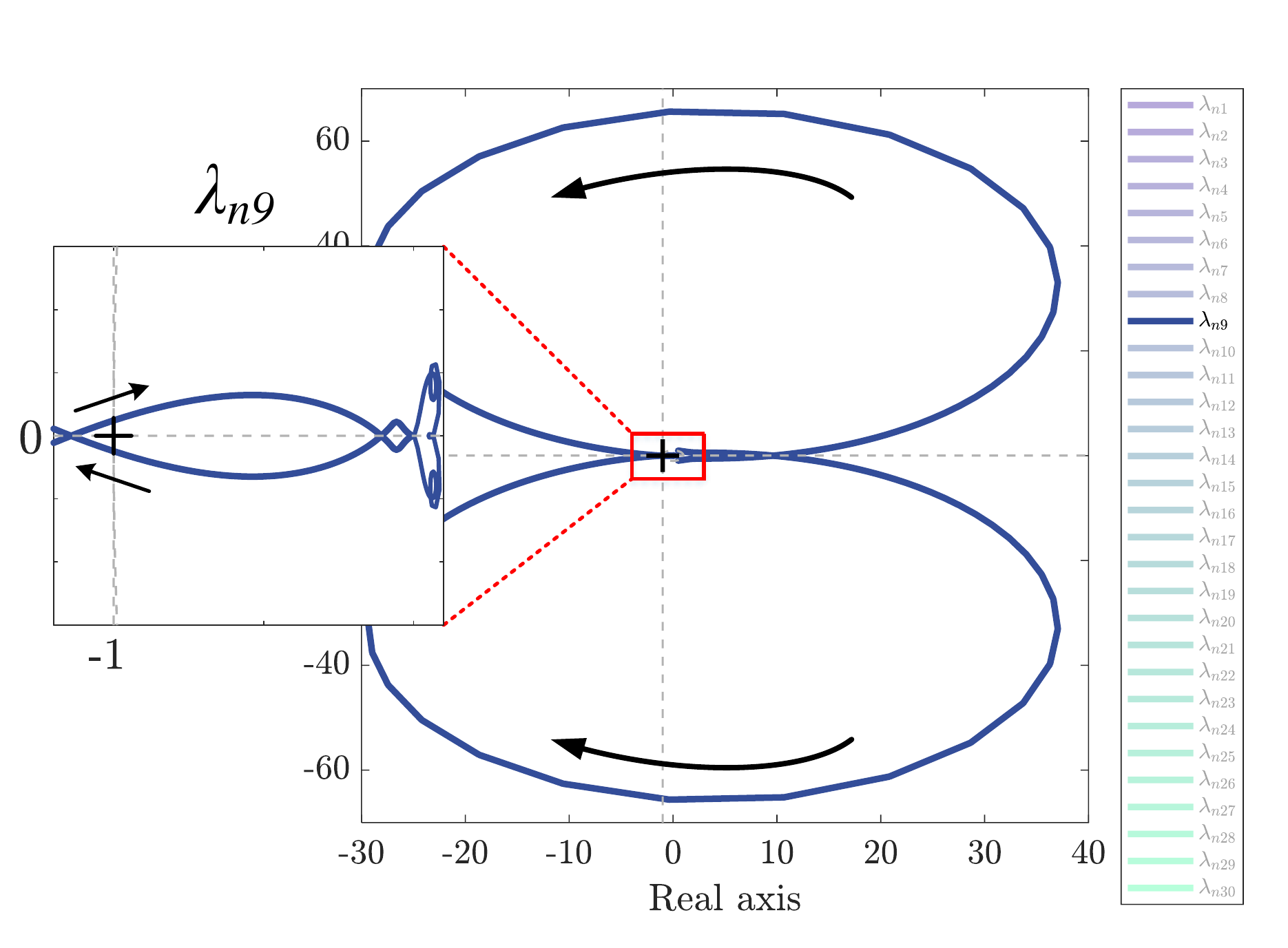}\label{subfig:T_nyqv}}
    
    \caption{Stability assessment case study II(b) GO1. (a) $\mathbf{Z_N}$ \& $\mathbf{Y_S}$, (b) Time domain simulation. (c) Eigenvalue analysis. (d) GNC.}
    \label{fig:T_ssassva}
    \end{figure*}

    In the GO1, the network is built according to Section III (i.e., passive components in $\mathbf{Z_N}$ and external elements to avoid interaction with resonances of the system in $\mathbf{Y_S}$ as displayed in Fig~\ref{subfig:T_toib}. According to~\cite{Abu-hashim1999TestSimulation}, VSC1 is connected to bus 3 through a transformer; therefore, a new bus, 15, is added for this purpose.
    
\begin{figure}
    \centering
    {\includegraphics[width=0.5\linewidth]{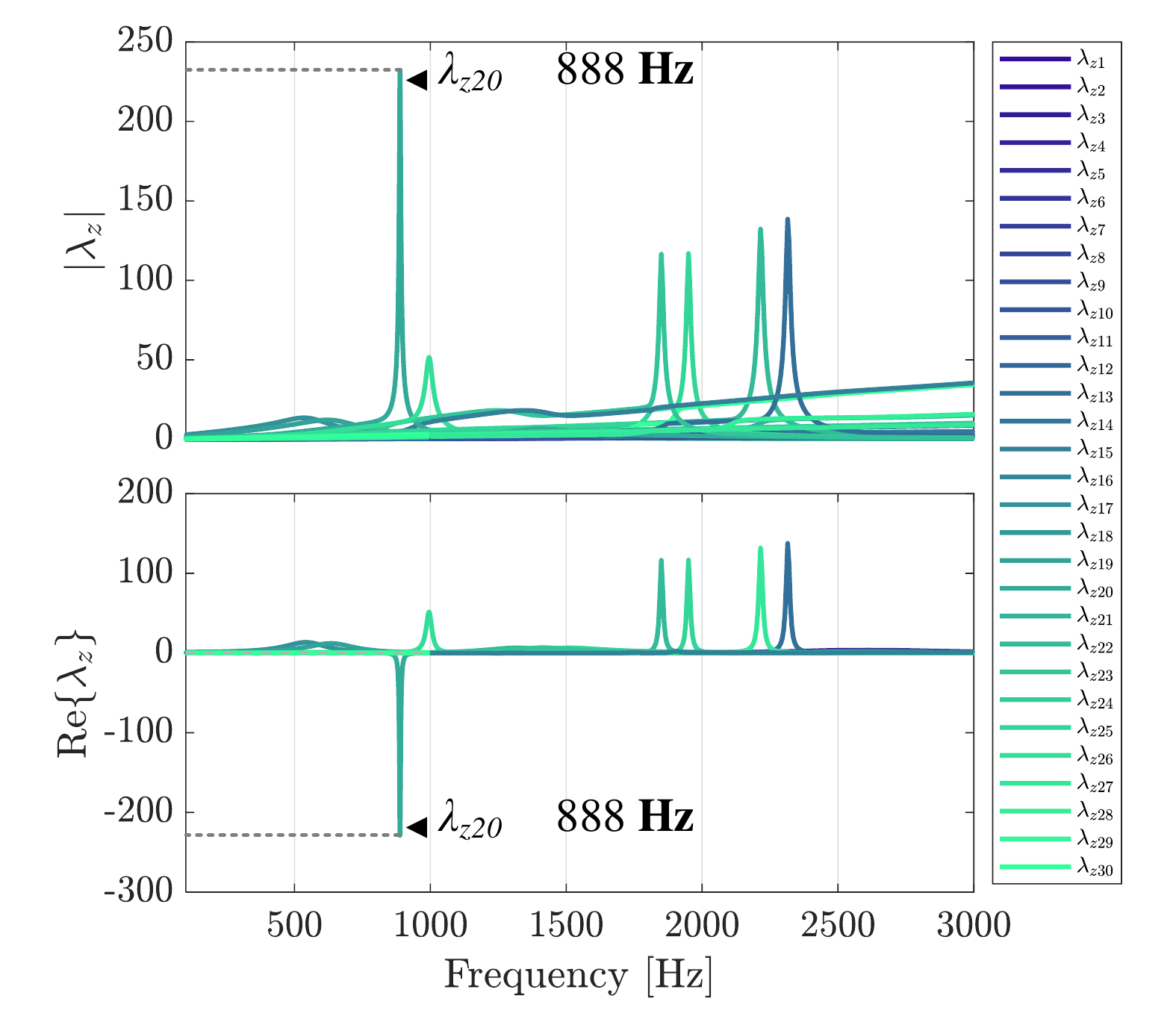}}
    \caption{Stability assessment case study II(b) GO1, PMD stability criterion.}
    \label{fig:T_fdv}
    \end{figure}

    In the time domain simulation, the non-linear Simulink model becomes unstable when $T_{fd}$ = 0.6 $T_{sw}$ of VSC1 as displayed in Fig~\ref{subfig:T_tdv}. The oscillatory unstable resonance is graphically determined at approximately 885 Hz. The instability is verified with the pole-zero plot of the linear $\mathbf{Z_T}(s)$ in the $s$-domain in Fig~\ref{subfig:T_pzv}, where there is a pair of poles in the RHP with an oscillating frequency of 888 Hz. In the frequency domain, the instability is corroborated with the GNC where $\lambda_{n9}$ curve encircles the critical point in clockwise direction as shown in Fig~\ref{subfig:T_nyqv}. Similarly, the proposed PMD stability criterion predicts the stability in Fig~\ref{fig:T_fdv} at the point where one of the eigenvalues of $\mathbf{Z_T}(j\omega)$, $\lambda_{z20}$, has a peak at 888 Hz in magnitude and the real part is negative.
    
    \paragraph{Grouping option 2}
    
\begin{figure*}[htbp]
    \centering
    \subfigure[]{\includegraphics[width=0.45\linewidth]{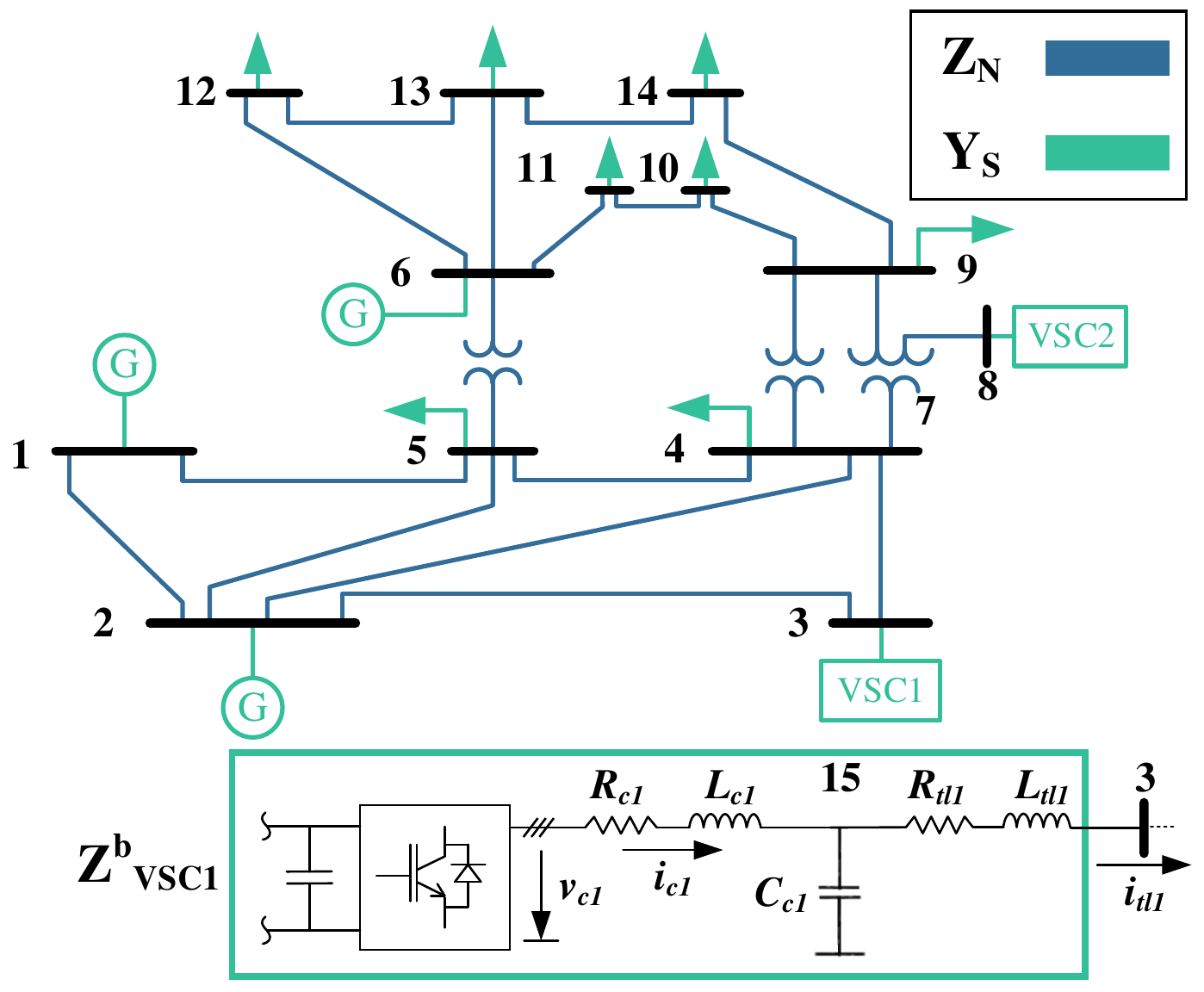}\label{subfig:T_toiib}}
    \subfigure[]{\includegraphics[width=0.48\linewidth]{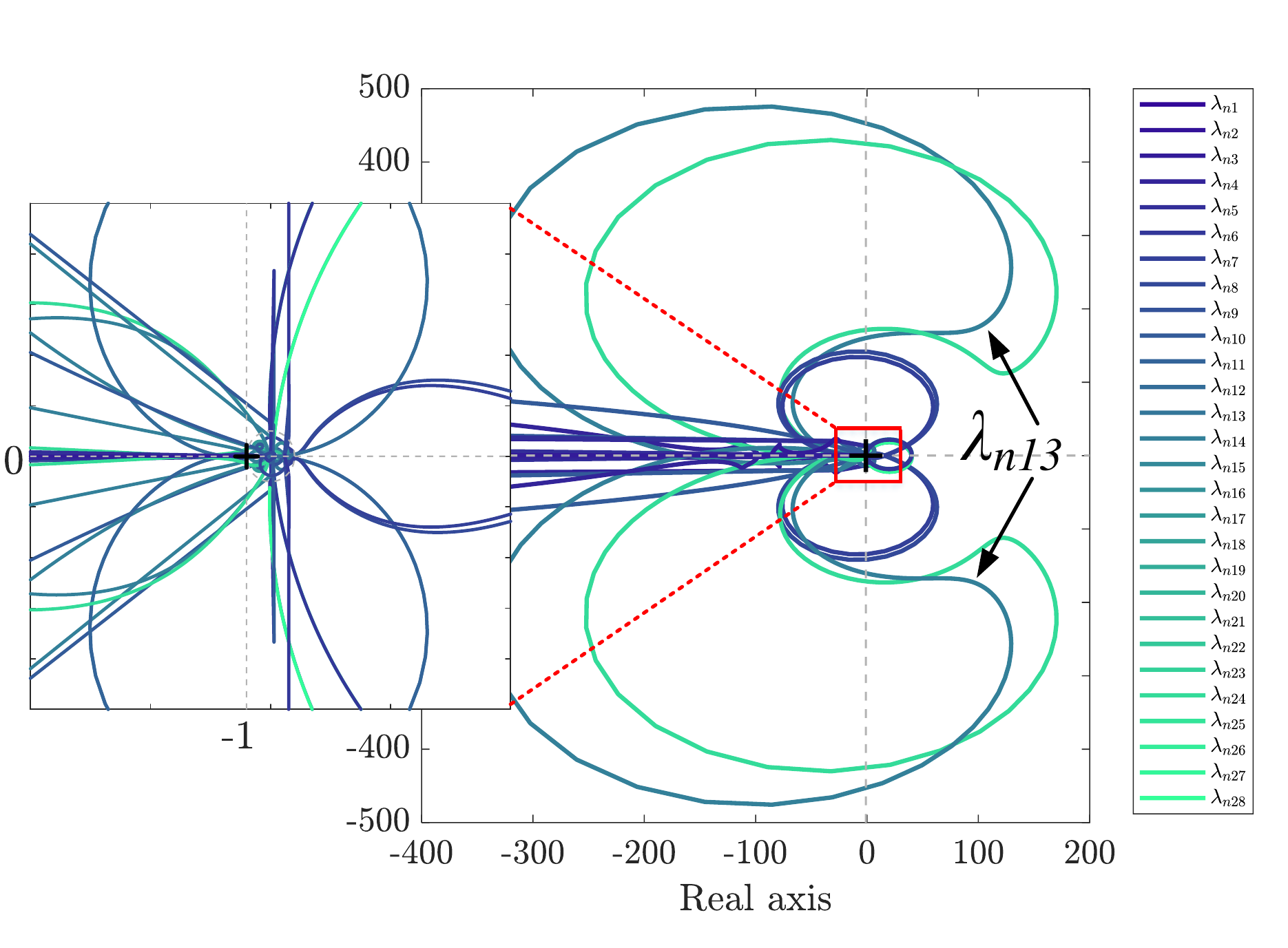}\label{subfig:T_nyqvia}}
    \subfigure[]{\includegraphics[width=0.85\linewidth]{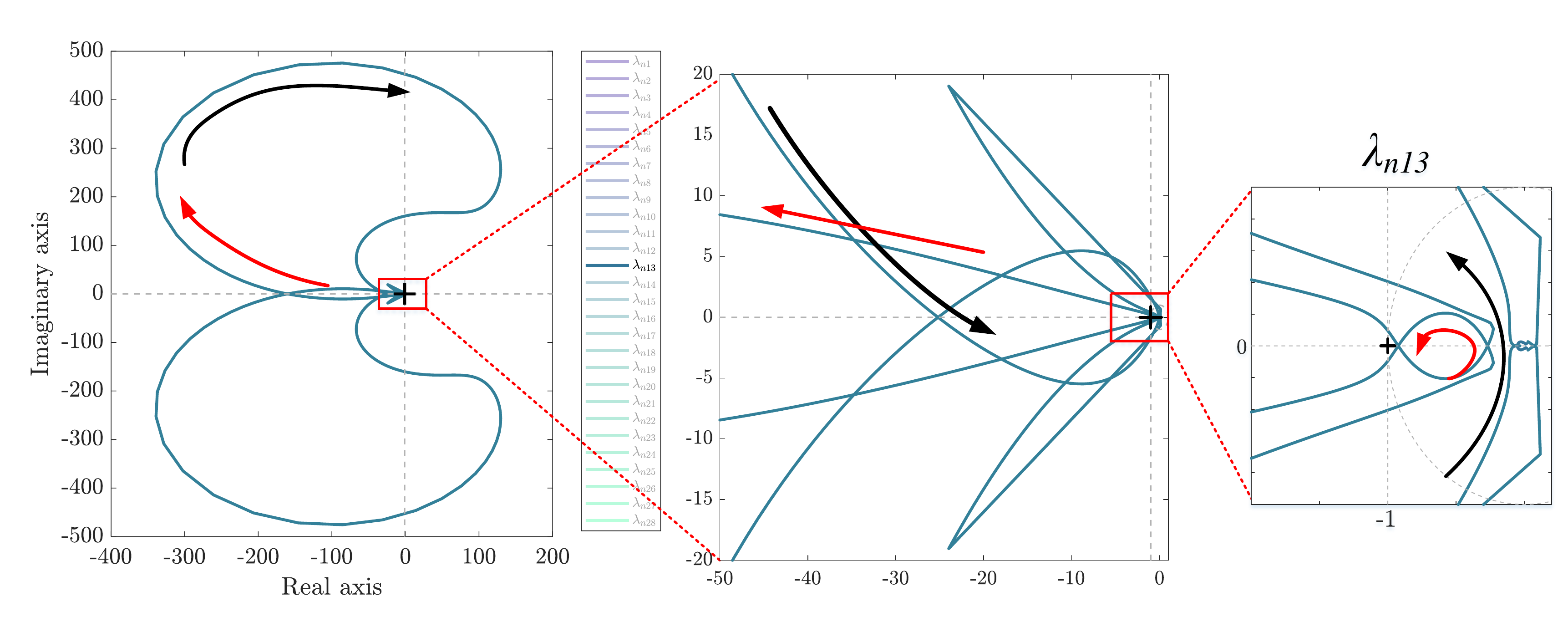}\label{subfig:T_nyqvib}}
    \caption{Stability assessment case study II(b) GO2. (a) $\mathbf{Z_N}$ \& $\mathbf{Y_S}$, (b) GNC all eigenvalues. (c) GNC $\lambda_{n13}$.} 
    \label{fig:T_ssassvia}
    \end{figure*}

    In the GO2, the system components are associated in order to cause RHP in the open-loop $\mathbf{L}$ as shown in Fig~\ref{subfig:T_toiib}. VSC1, the filter $\mathbf{Z_{cc1}}$, and the transformer $\mathbf{Y_{tl1}}$, connected at bus bar 15, are merged into $\mathbf{Y^b_{vsc1}}$ = $(\mathbf{Z^b_{vsc1})^{-1}}$  = $[\mathbf{Z_{vsc1}}//(\mathbf{Z_{cc1}} + \mathbf{ Y_{tl1}})]^{-1}$. The bus bar 15 is no longer needed (i.e., $\mathbf{Y^b_{vsc1}}$ is directly connected to bus 3); therefore, the $\mathbf{Z_T}$ order is reduced from 30 to 28.
    
\begin{figure}
    \centering
    {\includegraphics[width=0.5\linewidth]{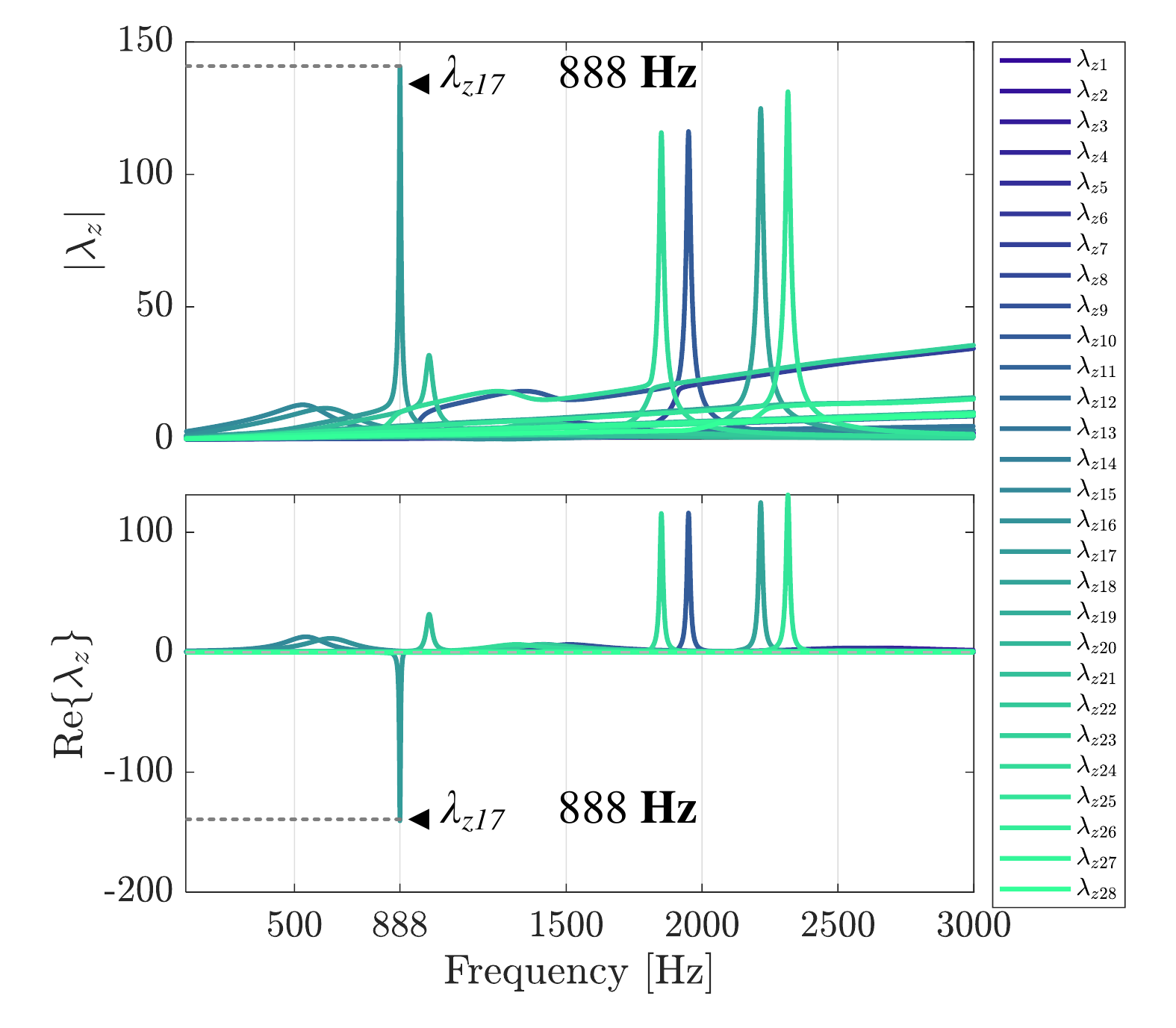}}
    \caption{Stability assessment case study II(b) GO2, PMD stability criterion.}
    \label{fig:T_fdvi}
    \end{figure}

    The stability of $\mathbf{Z_T}$ for GO2 is assessed in the frequency domain with the GNC and PMD stability criterion. In Fig~\ref{subfig:T_nyqvia}, the large number of Nyquist curves makes it difficult to address the stability straight away. For example, the Nyquist curve, $\lambda_{n13}$, encircles the critical point after making many turns close to the critical point, but in counterclockwise direction as displayed in Fig~\ref{subfig:T_nyqvib}. The GNC in this case  fails to predict the stability. On the other hand, in Fig~\ref{fig:T_fdvi}, the PMD stability criterion easily predicts the stability at the point where there is a peak at 888 Hz in the magnitude of one the eigenvalues magnitude of $\mathbf{Z_T}(j\omega)$, $\lambda_{z17}$, and the real part of it is negative.
    
    These are the findings of applying the PMD stability criterion to large and complex networks.
     
    \begin{itemize}[leftmargin=*]
    
        \item The instability and the resonance mode frequency can be easily identified through eigenvalue $\abs{\lambda_{zi}(j\omega)}$ and $\textrm{Re}{\{\lambda_{zi}(j\omega)}\}$ curves for high-order $\mathbf{Z_T}(j\omega)$ matrices compared to GNC. They provide a simpler visualisation of the physical interpretation of oscillatory modes and instabilities.
        \item The PMD stability criterion works with the complete nodal admittance matrix $\mathbf{Y_T}(j\omega)$, this is the reason why it is not affected by the aggregation of system elements in comparison with the GNC, which only studies the open-loop $\mathbf{L}$.
        \item The fact of working in the frequency domain allows to use black-box models or models based in measurements which can be added to $\mathbf{Y_T}(j\omega)$ in the frequency domain.
        \item Computation effort has also been reduced, because for the PMD stability criterion it is only needed to plot the frequency range of interest. On the other hand, the GNC requires a higher computation effort (e.g., store larger matrix arrays in memory  and longer processing time)  by plotting a wider frequency range  $\{ \omega \in \mathbb{R} \mid -\infty < \omega < +\infty \}$.
    \end{itemize}

    
\section{Conclusion}

    The present work contributes with the PMD stability criterion to address the drawbacks found in commonly used stability criteria to study multi-infeed grid-connected VSCs. The PMD stability criterion can only be used in power systems which satisfy the condition $\sigma_0 << \omega_0$, which is the main concern in power system stability studies.
    
    The article starts by testing the effectiveness of the PMD stability criterion on a three-bus simple network with two grid-connected VSCs. The complexity of the testing networks is then increased from a seven-bus network with three grid-connected VSCs to a 14 bus system with two VSCs connected. The networks are linearized in Matlab/Simulink according to state-space and impedance-based modelling techniques. Finally, the stability assessment results are compared with those from the eigenvalue analysis and the GNC criterion.
    
    This criterion has proven to be a powerful tool compared to eigenvalue analysis and the GNC, because it has the following relevant characteristics for a stability criterion: (i) frequency characterization of the closed-loop unstable oscillatory modes; (ii) does not require detailed information (e.g., use of measurements from black-box models); (iii) not sensitive to associations of systems elements; (iv) simple to evaluate and less computation effort (i.e., short computation time and memory usage); and (v) visually friendly and physical interpretation of results. 
    
    The PMD stability criterion has been successfully tested when addressing the stability of large power system with high-penetration of power electronics converters. However, some future work was identified during the present work such as
    
    \begin{itemize}
        \item working with experimental measurements and black-box models;
        \item examining the possibility to define stability margins;
        \item determining the influence of the operation point in linearized models over the stability;
        \item studying and quantifying the power system nodes contribution over the unstable resonance modes;
        \item and identifying how the network elements of the system affect the instability.
    \end{itemize}

    
\section{Appendix. VSC modelling}
    
    The grid-connected VSC small-signal model displayed in Fig.~\ref{subfig:T_cs} can be formulated as: 
    
    \subsubsection{Outer loop controller}
    
    The outer loop controls active power with the q-component ($\Delta i^c_{ref-q}$) and reactive power with the d-component ($\Delta i^c_{ref-d}$).  The current references can be defined as: $ \Delta i^c_{ref-q}$=$-F_{olp} \Delta p^c $  and $ \Delta i^c_{ref-d}$=$-F_{olq} \Delta q^c $, where $ F_{olp}=k_{p-olp}+k_{i-olp}/s $ and  $ F_{olq}=k_{p-olq}+k_{i-olq}/s $. The small signal active power ($\Delta p^c$) and reactive power ($\Delta q^c$) are the following, \begin{eqnarray}
    \Delta p^c = \dfrac{3}{2}[(i^c_{c0-q} \Delta v^c_q + v^c_{0-q} \Delta i^c_{c-q} + i^c_{c0-d} \Delta v^c_d + v^c_{0-d} \Delta i^c_{c-d})]\\  \label{eq:apower}
    \Delta q^c = \dfrac{3}{2}[(i^c_{c0-d} \Delta v^c_q - v^c_{0-d} \Delta i^c_{c-q} - i^c_{c0-q} \Delta v^c_d + v^c_{0-q} \Delta i^c_{c-d})], \label{eq:rpower}
    \end{eqnarray}
    
    \noindent where $v^c_{0-q}, v^c_{0-d}, i^c_{c0-q}, i^c_{c0-d} $ are voltages and currents at the linearization point.
    
    \subsubsection{Inner Loop Controller}
    
    The small-signal voltage modulated by the converter ($\Delta v^c_{c-ref}$) and the voltage measurement ($\Delta V$) can be expressed like \begin{eqnarray}
    \begin{bmatrix}\Delta v^c_{ref-q} \\ \Delta v^c_{ref-d} \end{bmatrix} = \begin{bmatrix}\Delta v^c_{h-q} \\ \Delta v^c_{h-d} \end{bmatrix}  -F_{il} \begin{bmatrix}\Delta i^c_{ref-q} \\ \Delta i^c_{ref-d} \end{bmatrix} + \begin{bmatrix} F_{il} & -\omega L_c\\ \omega L_c & F_{il} \end{bmatrix} \begin{bmatrix} \Delta i^c_{c-q} \\ \Delta i^c_{c-d} \end{bmatrix}
    \end{eqnarray}
    
    \noindent and \begin{eqnarray}
    \begin{bmatrix}\Delta v_q \\ \Delta v_d \end{bmatrix} = \begin{bmatrix}\Delta v_{c-q} \\ 
    \Delta v_{c-d} \end{bmatrix} + \begin{bmatrix} R_c+L_c s & \omega L_c \\ -\omega L_c & R_c+L_c s \end{bmatrix}  \begin{bmatrix} \Delta i_{c-q} \\ \Delta i_{c-d} \end{bmatrix},
    \end{eqnarray}
    
    \noindent where $ [\Delta v^c_{h-q} \enspace \Delta v^c_{h-d}]^T$ = $H_v [\Delta v^c_q \enspace \Delta v^c_d]^T $ and $ [\Delta v^c_{c-q} \enspace \Delta v^c_{c-d}]^T $ = $F_D [\Delta v^c_{ref-q} \enspace \Delta v^c_{ref-d}] $. $H_v$ and $F_D$ are the first-order feed-forward filter and the fifth-order Padé approximant delay, respectively. The PI controller $ F_{il}=k_{p-il}+k_{i-il} /s $ and its gains are $ k_{p-il}=\dfrac{L_c}{\tau _{il}}$  and  $ k_{i-il}=\dfrac{R_c}{\tau _{il}}$ \cite{Harnefors1998Model-basedMethod}. 
    
    \subsubsection{Phase-locked loop}
    
    The small-signal angle can be obtained with the following expression $\Delta \theta = - F_{pll} \Delta v^c_d$, where $ F_{pll}=k_{p-pll}+k_{i-pll}/s $. The rotation from a local to a global reference ($\mathbf{T}_{qd}^{-1}$)  and vice-versa ($\mathbf{T}^v_{qd}$) can be modelled as \begin{eqnarray}
    \begin{bmatrix}
    \Delta v^c_{c-q}\\
    \Delta v^c_{c-d}\\
    \end{bmatrix} =
    \begin{bmatrix}
    \cos{\Delta \theta_0}  & -\sin{\Delta \theta_0}\\
    \sin{\Delta \theta_0} & \cos{\Delta \theta_0} 
    \end{bmatrix}
    \begin{bmatrix}
    \Delta v_{c-q}\\
   \Delta v_{c-d}\\
    \end{bmatrix}
    + 
    \begin{bmatrix}
    -\Delta v_{c-q0} \sin{\Delta \theta_0} - \Delta v_{c-d0} \cos{\Delta \theta_0} \\
    \Delta v_{c-q0} \cos{\Delta \theta_0} - \Delta v_{c-d0} \sin{\Delta \theta_0}
    \end{bmatrix}
    \Delta \theta
    \label{eq:vtransformation}
    \end{eqnarray}
    
    \noindent and \begin{eqnarray}
    \begin{bmatrix}
    \Delta v_q\\
    \Delta v_d\\
    \end{bmatrix} =
    \begin{bmatrix}
    \cos{\Delta \theta_0}  & \sin{\Delta \theta_0}\\
    -\sin{\Delta \theta_0} & \cos{\Delta \theta_0} 
    \end{bmatrix}
    \begin{bmatrix}
    \Delta v^c_{q} \\
   \Delta v^c_{d} \\
    \end{bmatrix}
    + 
    \begin{bmatrix}
    -\Delta v^c_{q0} \sin{\Delta \theta_0} + \Delta v^c_{d0} \cos{\Delta \theta_0}\\
    -\Delta v^c_{q0} \cos{\Delta \theta_0} - \Delta v^c_{d0} \sin{\Delta \theta_0}
    \end{bmatrix}
    \Delta \theta ,
    \label{eq:ivtransformation}
    \end{eqnarray}
    
    \noindent where $\Delta \theta = \Delta \theta_0 - \Delta \theta_0^c$ (i.e., difference between the small-signal angle in the global reference and the angle in the local reference)~\cite{Rygg2017AAnalysis}. The expression in~\eqref{eq:vtransformation} is also used to change the converter current ($\mathbf{T}^i_{qd}$) from a global to a local reference. 

\bibliographystyle{IEEEtran}

\bibliography{./References/references.bib}

\end{document}